# A Physics-Informed Framework for PID Tuning of Chemical Processes Using Large Language Model Agents

Zhoupeng Shou[1], Xiaodong Hong[1,2,*], Congjing Ren[1,3], Jingdai Wang[1], Yongrong Yang[1], Zuwei Liao[1,*]
[1]State Key Laboratory of Chemical Engineering, Department of Chemical and Biological Engineering, Zhejiang University, Hangzhou, China; [2]Engineering Research Center of Functional Materials Intelligent Manufacturing of Zhejiang Province, ZJU-Hangzhou Global Scientific and Technological Innovation Center, Hangzhou, China; [3]Ningbo Innovation Center, Zhejiang University, Zhejiang, China
Email: hongxiaodong@zju.edu.cn; liaozw@zju.edu.cn

**Abstract:** PID tuning for chemical processes commonly relies on identified process models, whereas plant engineers often retune loops iteratively by observing responses, diagnosing deficiencies, adjusting gains, and validating the result. This work formalizes this engineer-like workflow in a language-model-assisted PID tuning framework applicable to both large and small language models (LLMs/SLMs). Hosted LLMs receive closed-loop response features, control-engineering diagnoses, tuning preferences, and internal model control (IMC)-based demonstrations to generate and iteratively correct PID gains under common acceptance criteria. For local deployment, Qwen3-0.6B is adapted through supervised fine-tuning (SFT) with simulation-verified IMC targets and physics-informed group relative policy optimization (PI-GRPO) with non-compensable stability and performance rewards. On 100 first-order plus dead time (FOPDT) and 100 second-order plus dead time (SOPDT) test cases, hosted LLMs (DeepSeek-V4-Flash and Qwen3.7-Plus) achieve final success rates of 75–89% and 77–79%, respectively. As for Qwen3-0.6B, supervised fine-tuning raises first-recommendation success to 86.5%, and PI-GRPO further increases it to 94.0%, primarily improving first-attempt reliability and stability margins.

## 1. Introduction

Proportional–integral–derivative (PID) control remains the predominant regulatory-control strategy in chemical plants[1,2]. PID tuning is commonly approached through two complementary routes. Model-based methods first identify process dynamics and then calculate controller gains using rules such as internal model control (IMC), whereas plant engineers often retune existing loops empirically by inspecting the closed-loop response, diagnosing deficiencies such as excessive overshoot or slow settling, adjusting the gains, and validating the resulting behavior. Classical tuning rules[3–5] and numerical optimization[6] provide useful baselines, but they rely on fixed process models, predefined objectives, or repeated parameter searches and do not inherently provide response interpretation or safety screening[7]. Engineer-led adjustment is more flexible but relies heavily on individual experience and lacks a systematic and transferable decision mechanism. These limitations have motivated data-driven adaptive tuning, stability-aware learning, and, more recently, language-model-assisted engineering decision making.

Reinforcement learning (RL) provides a data-driven approach for adaptive PID tuning. Autonomous tuning has been formulated by representing response characteristics as the state, gain adjustments as the action, and closed-loop performance as the reward[8]. Hybrid approaches preserve PID as the deployable controller while using deep RL to identify improved gains[9], and actor–critic algorithms provide flexible tools for continuous process-control actions[10]. Practical implementation studies have clarified the additional challenges of learning in continuous processes rather than idealized benchmark environments[11]. Offline learning has further been used for an ethylene-cracking-furnace feeding strategy, avoiding exploratory interaction with the live plant[12]. These developments improve adaptability, yet they also sharpen the deployment question: improvement in a scalar reward does not ensure that the resulting controller is safe or robust for a chemical-process control loop.

Recent learning-enabled control research has therefore placed greater emphasis on explicit physical and stability information. Safe model-based RL has incorporated Lyapunov and barrier-function conditions to address state and input constraints[13], while physics-informed RL has embedded stability knowledge into policy optimization for nonlinear systems[14]. Recent tutorial work identifies safety guarantees, uncertainty, model knowledge, and scalability as continuing obstacles to RL-based nonlinear control[15]. Supervisory safety mechanisms and backup controllers provide one route for rejecting unsafe learned actions. Cui et al.[16] developed a shielded economic RL framework in which a supervisory layer accepts an RL action only when prescribed Lyapunov-stability and safety conditions are satisfied; otherwise, it invokes a Lyapunov-based economic model-predictive-control (MPC) or a stabilizing fallback controller to preserve safe operation. Zhu et al.[17] proposed an RL method for stochastic nonlinear systems that constructs a neural stochastic control Lyapunov function and a Sontag-type policy, establishing learning convergence and asymptotic stability in probability under both continuous and sample-and-hold implementations. These developments establish a useful principle for PID tuning: performance improvement should be accompanied by verifiable closed-loop behavior and an engineering mechanism for rejecting unacceptable decisions.

Large language models (LLMs) offer a complementary capability that is particularly suited to response-driven, engineer-like tuning. They can organize process characteristics, response observations, control requirements, engineering constraints, and user preferences into a structured decision context. SmartControl[18] translates natural-language requirements into PID design targets and couples this interaction layer with numerical optimization. Agentic PID workflows have also shown that language-model reasoning can participate in real-time parameter adjustment of a servo motor when safe ranges and feedback are imposed[19]. LLM tuning assistants and controller-design agents have been applied to cascade structures, input–output pairing, and automated tuning workflows[20,21].

A recent water-tank study further compared direct LLM actuation with hybrid LLM–PID supervision[22]. Direct language-model control failed to regulate the process, whereas supervisory gain adjustment improved nominal transient performance but produced irregular updates under sustained measurement drift. Furthermore, fine-tuned language models have been investigated as sequence-to-parameter estimators for adaptive NMPC, mapping closed-loop trajectory summaries to uncertain process parameters and providing learned initializations for subsequent refinement[23]. Related work in process systems engineering has also shown that LLM agents can interpret engineering descriptions, invoke computational tools, generate simulator inputs, and verify executable process representations[24–26]. Together, these studies support the use of LLMs as high-level engineering assistants, but not as unconstrained low-level controllers.

For chemical-process applications, fluent parameter recommendations are insufficient unless they are grounded in process evidence and evaluated against control constraints. Hybrid-AI frameworks emphasize the integration of learning with domain knowledge, explicit constraints, model validation, and engineering governance[27,28]. Few-shot prompting provides a practical mechanism for supplying representative tuning cases without retraining a large language model[29], while structured retrieval systems such as ChatP&ID[30] show that engineering-oriented LLM interaction becomes more transparent and reliable when grounded in organized process information. In PID tuning, closed-loop response features, engineering diagnoses, process dead time, controller limits, and validation outcomes can provide this grounding. Static prompting alone, however, cannot fully represent the current response deficiency or incorporate the measured consequence of a previous gain recommendation. The tuning process should instead follow an iterative observe–diagnose–adjust–validate cycle, analogous to engineering commissioning practice, while retaining explicit analytical and closed-loop checks to prevent unconstrained trial and error.

Deployment presents a second limitation. Hosted LLMs introduce inference latency, recurring cost, data-governance concerns, and output variability. Recent work has therefore begun to investigate small language models (SLMs) for dynamic-system control. Locally deployed Qwen2.5 models have been used to generate model predictive control cost functions from natural-language objectives, with self-assessment substantially improving control success across several simulated systems[31]. However, the pronounced performance difference between the tested 7B and 3B models also shows that local execution alone does not ensure reliable control behavior. Post-training methods have demonstrated that language-model behavior can be adapted through supervised fine-tuning, preference optimization, and reinforcement learning with verifiable rewards[32–35]. SLMs are therefore attractive for specialized and resource-constrained deployment when domain capability can be introduced through targeted adaptation[36]. For PID tuning, however, direct regression of controller gains may reproduce training labels without learning how stability, response quality, and engineering constraints should influence parameter selection.

Despite these advances, existing studies address only individual components of language-model-assisted control. Hosted LLMs have been used for supervisory PID adjustment, fine-tuned language models for NMPC parameter estimation, and SLMs for MPC cost-function generation. However, a unified PID-tuning formulation that formalizes the engineer-like cycle of response observation, diagnosis, gain adjustment, and closed-loop validation has not yet been established. In particular, existing studies do not combine this iterative reasoning process with explicit controller-acceptance criteria and physics-informed adaptation of a lightweight local model using verifiable stability and performance rewards.

To address this problem, this study develops a language model-assisted PID tuning framework for delayed chemical processes. The study proceeds in two stages. First, hosted LLMs act as virtual control experts by following an observe–diagnose–adjust–validate cycle. Closed-loop response indicators are translated into engineering diagnoses and combined with tuning preferences and feedback from previous recommendations. The resulting dynamic prompts support targeted gain correction, while common closed-loop acceptance criteria determine whether each recommendation is retained or rejected. Second, Qwen3-0.6B is adapted to the same tuning task for resource-efficient local deployment. Supervised fine-tuning establishes an engineering-feasible mapping from process descriptions and control objectives to PID gains, while physics-informed group relative policy optimization further refines the policy using stability- and performance-aware rewards. The resulting framework supports both hosted-model reasoning and lightweight, verifiable PID tuning under a common validation logic.

# 2. Methodology

## 2.1 Overall Framework

This study develops a response-driven framework for PID tuning in chemical processes, in which language models act as virtual control experts within a closed-loop validation workflow. Rather than treating the language model as an unconstrained parameter generator, the framework combines response analysis, engineering knowledge, model inference, and controller validation. As shown in Figure 1, it consists of two stages: dynamic-prompt-based tuning with hosted LLMs and physics-informed adaptation of a lightweight SLM.

In the first stage, closed-loop response data are converted into performance indicators and control-engineering diagnoses. These features, together with the current PID gains, process characteristics, tuning preferences, and representative examples, are organized into a structured prompt. The LLM then proposes candidate gains, which are evaluated against the prescribed closed-loop performance criteria. Unresolved response deficiencies are returned as feedback for subsequent tuning rounds.

In the second stage, Qwen3-0.6B is adapted to the same PID-tuning task for lightweight deployment. Supervised fine-tuning (SFT) establishes an engineering-feasible initial policy and a stable output format, while Physics-informed group relative policy optimization (PI-GRPO) further refines the policy using verifiable stability and closed-loop performance rewards. The resulting framework connects language-model reasoning with executable and validated PID tuning decisions.

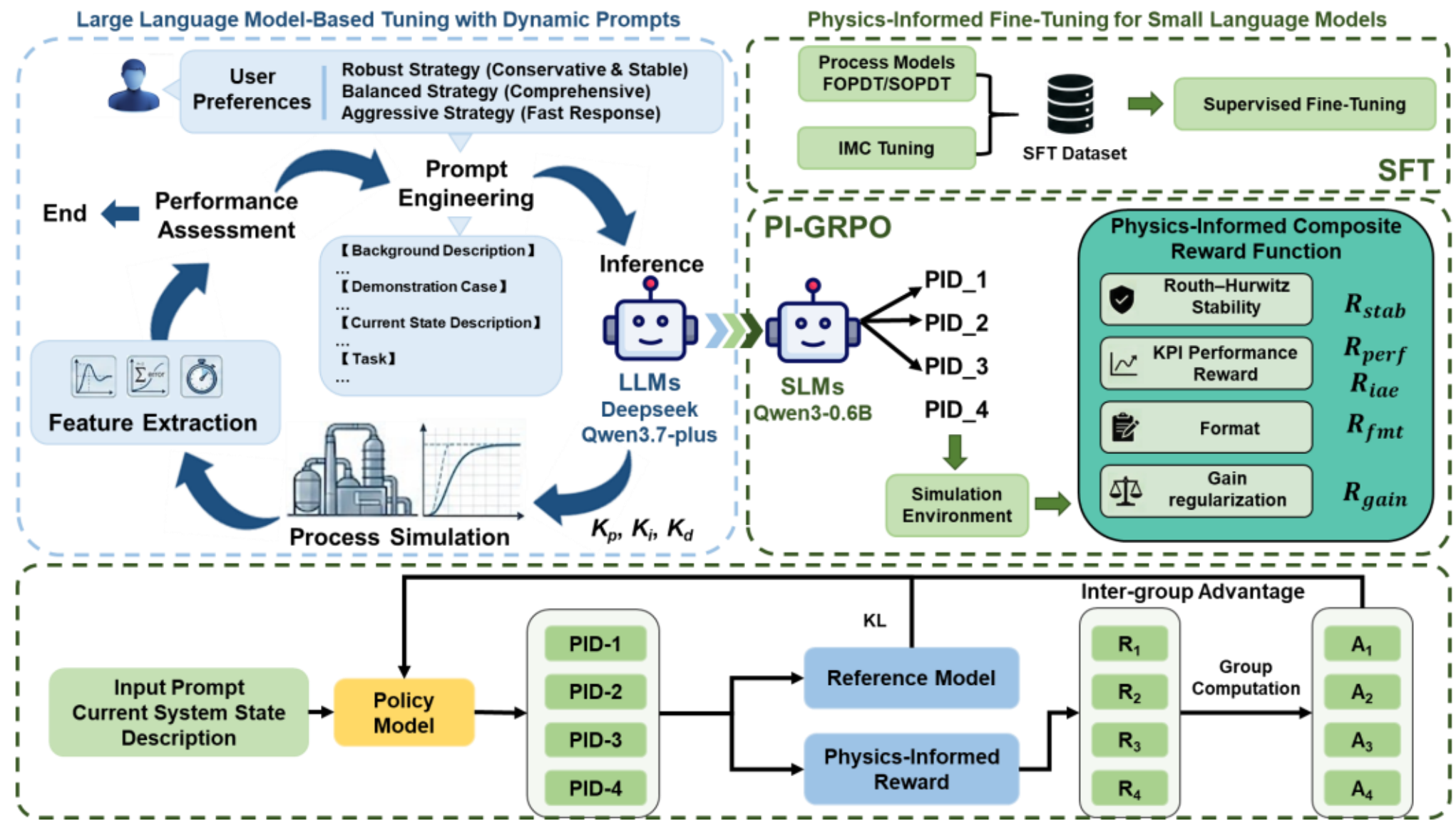

**Figure 1**. Language model-assisted PID tuning framework and physics-informed adaptation of a lightweight small language model.

## 2.2 Language Model-Based Tuning with Dynamic Prompts

### 2.2.1 Closed-Loop Tuning Architecture

The feature-extraction module characterizes each closed-loop response using the integral absolute error (IAE) and seven complementary dynamic features: overshoot, the number of significant setpoint crossings, attenuation ratio, oscillation period, post-dead-time rise time to 63.2% of the setpoint, settling time, and steady-state error. Detailed definitions are provided in Supporting Information S1.

The performance-assessment module determines whether a candidate controller satisfies the prescribed acceptance criteria. A response must remain finite, enter and remain within the ±5% settling band, and have an overshoot below 15% and a steady-state error below 1%. Overshoot ($M_p$) is evaluated from the largest sampled output, whereas the steady-state error ($e_{ss}$) is calculated from the mean of the final response window to reduce sensitivity to a single endpoint.

$$M_p = \max\left(0, \frac{\max_t (y_t - r)}{r}\right) \times 100\%, \;\; e_{ss} = \frac{|\mathrm{mean}(y_{(N-100:N)}) - r|}{|r|} \times 100\% \tag{1}$$

where $r$ is the setpoint, $y_t$ is the output at time $t$, and $N$ is the total number of samples.

If all acceptance criteria are satisfied, the tuning loop terminates and retains the candidate gains. Otherwise, the remaining deficiencies in overshoot, settling behavior, steady-state accuracy, and accumulated error are summarized in a feedback prompt. Each subsequent recommendation is therefore conditioned on the measured closed-loop consequence of the previous gain set rather than generated as an independent trial.

The prompt-construction module combines the current PID gains, extracted response features, performance assessment, previous feedback, process dead time, requested control style, and demonstration cases. The LLM is instructed to return a machine-readable PID triplet, which is parsed and evaluated using the same response metrics and acceptance criteria throughout the study. This common validation layer enables direct comparison across prompt variants and model families. In the present benchmark, closed-loop validation is implemented using the process simulation environment described in Supporting Information S2.

### 2.2.2 Dynamic Prompt: Expert Translation and Few-Shot Preference Injection

Numerical response indicators quantify control performance but do not always make their tuning implications explicit. We therefore translate them into concise control-engineering diagnoses. The semantic descriptions identify response characteristics such as insufficient damping, persistent error, slow settling, and excessive control aggressiveness, allowing the LLM to interpret the response as a controller-tuning problem rather than as an isolated numerical vector. The translation rules are deterministic and applied consistently across all models. For example, large overshoot is associated with inadequate damping or overly aggressive proportional or integral action, whereas a slow monotonic response is characterized as conservative. The original numerical indicators are retained, so the semantic descriptions supplement rather than replace the quantitative evidence.

PID tuning involves trade-offs among response speed, overshoot suppression, robustness, and control effort. To encode user preferences without modifying the LLM weights, few-shot examples are introduced as in-context guidance. The demonstration cases are generated using internal model control (IMC) rules. In these examples, the IMC filter coefficient $\lambda$ determines the speed–robustness trade-off: larger values produce more conservative tuning, whereas smaller values favor faster and more aggressive tracking. Demonstrations with larger $\lambda$ are selected when robustness, disturbance rejection, or smooth operation is preferred; demonstrations with smaller $\lambda$ are selected when rapid setpoint tracking is prioritized. In this way, qualitative control preferences are converted into contextual guidance for gain generation. Additional IMC-based demonstration cases are provided in Supporting Information S3.

Each initial prompt contains a demonstration block and a current-state block. The demonstration block provides examples matched to the process class and desired control style. The current-state block reports the present PID gains and IAE, process dead time, requested tuning style, numerical response features, and their engineering interpretation. In subsequent feedback rounds, the demonstration block is omitted, while the measured outcome of the previous gain set is retained to guide targeted correction. Prompt samples can be found in Supporting Information, while the prompt template is summarized as follows:

*[Background] Act as a control-engineering assistant for PID tuning.*
*[Demonstration Cases] Ten examples for the corresponding process class and control style.*
*[Current State] Current PID gains and IAE; observed dead time; desired control style; numerical response indicators and their engineering interpretation.*
*[Task] Return only P:<value>; I:<value>; D:<value>.*

### 2.3 Physics-Informed Fine-Tuning for Small Language Models

### 2.3.1 Supervised Fine-Tuning (SFT) and Group Relative Policy Optimization (GRPO)

An unadapted lightweight language model has limited ability to infer reliable PID gain adjustments from closed-loop response descriptions. Supervised fine-tuning (SFT) is therefore first used to establish a stable output format and an engineering-feasible initial policy. The SFT dataset contains 40,000 records, equally divided between first-order plus dead time (FOPDT) and second-order plus dead time (SOPDT) processes. All records use the balanced tuning style, with deterministic IMC controllers verified through closed-loop simulation as supervisory targets. Approximately half of the records are generated through a feedback-construction pathway, with up to three accepted intermediate feedback turns, exposing the model both to initial tuning requests and to corrective recommendations after an intermediate gain set fails to satisfy the acceptance criteria. Details of data generation and SFT training are provided in Supporting Information S4.

SFT provides a feasible initialization but does not directly optimize the controllers generated during inference. A physics-informed reinforcement-learning stage is therefore introduced, using group relative policy optimization (GRPO) [37]. For each process and control objective, the policy generates four candidate PID triplets, as illustrated in Figure 1. Each candidate is parsed, screened using analytical stability criteria, and evaluated through closed-loop validation. A physics-informed reward is then assigned according to stability, time-domain performance, IAE improvement, gain regularization, and output-format validity.

Within each four-candidate group, rewards are standardized to obtain relative advantages $A_1 - A_4$. Policy updates use a clipped probability-ratio objective together with a Kullback–Leibler (KL) penalty to the frozen SFT reference. A KL divergence weight is adjusted during training from the observed divergence, allowing the policy to improve while limiting departures from the supervised initialization.

Because multiple PID controllers may satisfy the same response requirements, the group-relative formulation favors better candidates for a given process without imposing a unique numerical target. An independent validation set is evaluated before training and at fixed intervals. A GRPO checkpoint is retained only if it outperforms the step-0 SFT baseline according to the predefined validation hierarchy. The optimization objective, adaptive KL rule, checkpoint-selection procedure, and training settings are provided in Supporting Information S4.

### 2.3.2 Control-Theoretic Reward Design

The physics-informed reward connects language-model outputs to verifiable closed-loop behavior. Its general form is written as:

$$R = f\left(K_p, K_i, K_d;\ \text{closed-loop response metrics}\right) \tag{3}$$

where $K_p$, $K_i$, and $K_d$ denote the proportional, integral, and derivative gains of PID controllers, respectively. The closed-loop response metrics include time-domain performance quantities such as IAE, overshoot, settling time, and steady-state error. The total reward follows a gated hierarchy:

$$R = \begin{cases} -1, & \text{invalid or unsafe} \\ -\frac{1}{2} + \frac{Q+1}{4}, & \text{safe but unsuccessful} \\ \frac{1}{2} + \frac{Q+1}{4}, & \text{safe and successful} \end{cases} \tag{4}$$

where the bounded quality score $Q$ is

$$Q = \text{clip}\left(0.35R_{stab} + 0.35R_{perf} + 0.15R_{iae} + 0.10R_{gain} + 0.05R_{fmt},\ -1,\ 1\right) \tag{5}$$

Here, $R_{stab}$, $R_{perf}$, $R_{iae}$, $R_{gain}$, and $R_{fmt}$ denote the stability reward, performance reward,

IAE-improvement reward, dimensionless-gain regularization, and format reward respectively. The weights are fixed and sum to one.

A candidate receives the fixed reward $R = -1$ if any of the following conditions is met: (1) the PID triplet cannot be parsed; (2) the gains violate $K_p > 0$, $K_i \geq 0$, or $K_d \geq 0$; (3) the closed-loop trajectory becomes non-finite or exceeds the prescribed output bound; (4) the Padé-based characteristic polynomial is classified as unstable, marginal, or indeterminate by the Routh–Hurwitz test; or (5) an exact-delay phase margin is available and is non-positive. A candidate that is Routh stable but has no detected unity-gain crossover remains eligible for the safe branches but receives $R_{stab} = 0$.

Among safe candidates, task success requires a settled response, an overshoot strictly below 15%, and a steady-state error strictly below 1%. Candidates that pass the safety gate but do not satisfy the task-success criteria are mapped to $[-0.5,0]$, whereas safe and successful candidates are mapped to $[0.5,1]$. A successful controller always outranks a safe controller that has not yet met the task criteria. Within each safe branch, the quality score $Q$ provides continuous differentiation among candidate controllers. This hierarchy makes safety and task completion non-compensable: favorable performance on a soft reward component cannot offset an unsafe stability assessment or replace the prescribed success criteria.

For each candidate PID parameter set, the first-order [1/1] Padé approximation is used only to derive the closed-loop characteristic polynomial for the Routh–Hurwitz stability test. The phase margin is evaluated separately from the exact delayed open-loop transfer function at all refined unity-gain crossover frequencies. Candidates that fail the Routh–Hurwitz test or exhibit a non-positive phase margin are assigned to the unstable branch of Eq. (4). Candidates passing this stability gate enter the stable branch, where $R_{stab}$ provides a graded reward according to the phase margin rather than a binary pass/fail score.

The performance reward $R_{perf}$ combines overshoot, settling time, and steady-state error, while the IAE-improvement reward $R_{iae}$ measures the reduction in accumulated tracking error relative to the preceding controller. The dimensionless-gain regularization $R_{gain}$ is a soft penalty that discourages disproportionately large proportional, integral, or derivative gains relative to deterministic IMC-derived reference scales. Normalization by the process gain and characteristic time makes this regularization comparable across FOPDT and SOPDT processes of different scales. It therefore suppresses unnecessarily aggressive parameterizations without imposing hard upper bounds or replacing the stability gate. Complete reward definitions, thresholds, and implementation equations are provided in Supporting Information S5.

## 3. Experimental Setup

### 3.1 Process Models and Benchmark Construction

The experiments use FOPDT and SOPDT models to represent processes with one or two dominant lags and a pure dead time. The same process definitions, closed-loop simulator, response metrics, and convergence criteria are used for hosted LLMs, Qwen3-0.6B variants, and the IMC baseline. The FOPDT and SOPDT models are defined as

$$G_{FOPDT}(s) = \frac{K \cdot e^{-\theta s}}{Ts+1}, G_{SOPDT}(s) = \frac{K \cdot e^{-\theta s}}{(\tau_1 s+1)(\tau_2 s+1)} \tag{6}$$

where $K$ is the process gain, $\theta$ is the dead time, $T$ is the FOPDT time constant, and $\tau_1$ and $\tau_2$ are the two SOPDT time constants.

For the FOPDT processes, $K$ is sampled from $[0.2, 0.9]$ and $T$ is sampled from $[100, 600]$ s. For SOPDT processes, $K$ is sampled from $[0.1, 3.0]$, the slower time constant from $[10, 100]$ s, and the ratio of the faster to the slower time constant from $[0.1, 1.0]$. The dead time $\theta$ varies from 1 to 200 s and is additionally constrained relative to the characteristic process time. This design covers low-delay, mediate-delay, delay-dominant, and strong-delay regimes.

The benchmark contains 100 FOPDT cases and 100 SOPDT cases. Continuous process parameters are generated using Latin hypercube sampling, while dead-time strata and controller-defect categories are assigned according to a predefined schedule. Independent random seeds are used for the FOPDT and SOPDT sets. All benchmark cases are excluded from the demonstration libraries and from the SFT and PI-GRPO training datasets. Each benchmark case begins from a controlled perturbation of its balanced IMC reference controller rather than from a common fixed gain set. Ten perturbation patterns represent excessive or insufficient proportional, integral, and derivative action, both individually and in combination. A case is retained only if its initial response remains finite but fails the acceptance criteria and its IAE is at least 1.5 times that of the balanced IMC reference. This procedure creates diverse and interpretable tuning tasks without using the final benchmark results during training.

Figure 2 summarizes the coverage of the benchmark cases. To compare dead-time effects across processes with different characteristic time scales, the dimensionless relative-delay ratios are defined as

$$\boldsymbol{\rho} = \begin{cases} \theta/\boldsymbol{T}, & \boldsymbol{for\ FOPDT} \\ \theta/(\tau_1 + \tau_2), & \boldsymbol{for\ SOPDT} \end{cases} \tag{7}$$

Figures 2(a) and 2(b) show that the sampled cases cover broad regions of the process-parameter spaces, while Figure 2(c) confirms representation across the prescribed relative-delay intervals. Figure 2(d) summarizes the controller-defect coverage. Each of the ten PID fault types contains ten FOPDT and ten SOPDT cases. For each process class, the severity distribution consists of 30 mild, 50 moderate, and 20 severe cases, giving 60 mild, 100 moderate, and 40 severe cases across the combined benchmark. The evaluation therefore covers both heterogeneous process dynamics and a structured range of controller-mismatch patterns. Detailed case definitions are provided in Supporting Information S6.

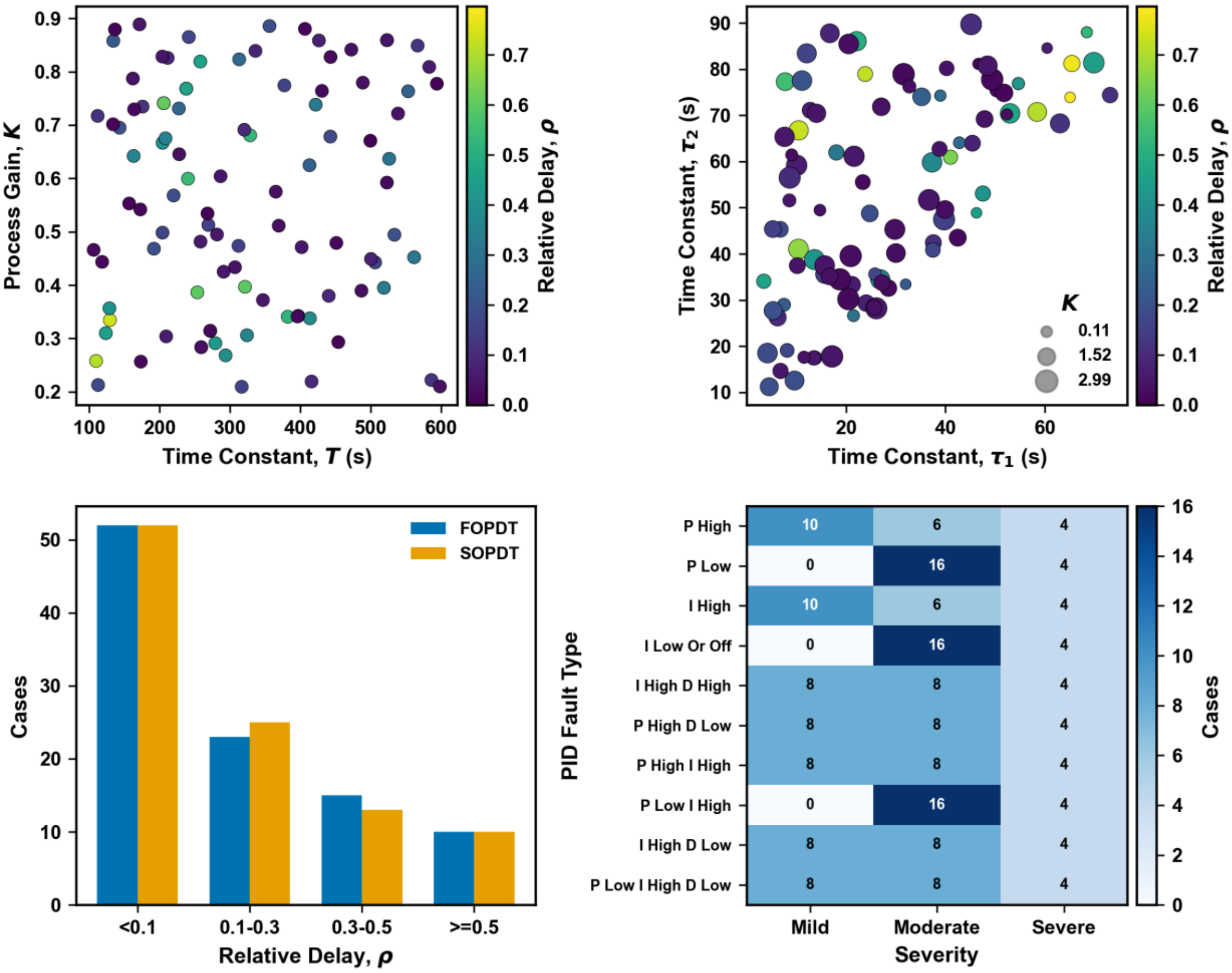

**Figure 2**. Coverage of the 200 benchmark cases. (a) Distribution of the 100 FOPDT cases in the process-gain and time-constant space, where color denotes the relative delay $\boldsymbol{\rho}$. (b) Distribution of the 100 SOPDT cases in the $\boldsymbol{\tau_1 - \tau_2}$ space, where color denotes the relative delay $\boldsymbol{\rho}$ and marker size represents the process gain $\boldsymbol{K}$. (c) Numbers of FOPDT and SOPDT cases within different relative-delay intervals. (d) Coverage of the prescribed PID fault types and severity levels across the combined evaluation set. The numerical annotations and color intensity indicate the number of cases in each category.

All simulations use a unit setpoint, a time horizon of 4000 s, and 40,001 uniformly spaced samples, corresponding to a sampling interval of 0.1 s. The process-specific dead time is applied explicitly. A trajectory is treated as invalid if the process output, tracking error, or control signal becomes non-finite, or if the absolute process output exceeds 3. Tuning performance is summarized using the cumulative Pass@$k$ metric. Pass@$k$ is defined as the percentage of test cases for which at least one of the first $k$ PID recommendations satisfies all acceptance criteria. Pass@1 therefore measures first-recommendation success without feedback correction.

### 3.2 Language Model Selection and Adaptation

DeepSeek-V4-Flash and Qwen3.7-Plus, both accessed through their official application programming interfaces (APIs), are selected as representative cloud-hosted large language models (referred to as hosted LLMs). DeepSeek-V4-Flash employs a mixture-of-experts architecture with 284 billion total parameters and approximately 13 billion parameters activated per token. Qwen3.7-Plus provides an independent hosted baseline from a different model family. The two hosted LLMs are evaluated using identical prompts, feedback rules, output constraints, response metrics, and closed-loop acceptance criteria. Because the services may differ in architecture, model scale, training data, post-training procedures, and inference infrastructure, their comparison is treated as an empirical evaluation of end-to-end PID-tuning performance rather than as a controlled analysis of parameter count or model architecture.

Qwen3-0.6B is selected as the lightweight, locally deployable model because of its small memory footprint and low inference cost. Its base, supervised fine-tuned, and PI-GRPO-adapted variants are evaluated to determine whether reliable PID-tuning behavior can be acquired at a substantially smaller model scale. This adaptation procedure does not directly distill the responses of the hosted LLMs. Instead, control-domain knowledge is introduced through model-based supervision and physics-informed feedback. The comparison therefore evaluates the practical performance gap between hosted LLM services and a task-adapted lightweight local model, rather than providing a controlled comparison among model generations or architectures. Qwen3-0.6B training is conducted on a server equipped with five NVIDIA A100 GPUs, each with 80 GB of memory. Local inference requires only one GPU.

### 3.3 SFT and PI-GRPO Settings

Qwen3-0.6B is first trained by full-parameter SFT for up to three epochs using bf16 precision, a learning rate of $1\times10^{-5}$, and a maximum sequence length of 6144 tokens. Five percent of the 40000 records form a validation split, and the checkpoint with the best validation loss is restored. PI-GRPO then uses 4-bit NF4 Quantized low-rank adaptation (QLoRA), a LoRA rank of $r = 16$, a scaling parameter of $\alpha = 32$, and zero dropout. The learning rate is warmed up to $1\times10^{-5}$ over 50 steps and decayed using a cosine schedule to $1\times10^{-6}$ by step 800. At each optimization step, four online prompts are sampled on each of the five GPUs, and four candidate completions are generated for every prompt. The clipped update range is 0.2, and the adaptive KL coefficient starts at 0.10 with a target divergence of 0.05. Greedy validation is run before training and every 50 steps, with the step-0 SFT model retained unless a GRPO checkpoint improves the predefined selection hierarchy. Detailed settings are listed in Table S1 of Supporting Information Section S4.

## 4. Results and Discussion

### 4.1 Performance of Large Language Model-Assisted PID Tuning Framework

#### 4.1.1 FOPDT and SOPDT Systems

Four representative cases are selected from the 100 FOPDT and 100 SOPDT benchmark sets to illustrate the response-correction behavior of the language-model-based tuning workflow. C1 and C3 are delay-dominant processes, with relative-delay ratios of $\rho = 0.412$ and $\rho = 0.423$, respectively, whereas C2 and C4 are medium-relative-delay processes, with $\rho = 0.078$ and $\rho = 0.089$. Figure 3 compares the initial controller with the IMC baseline, DeepSeek-V4-Flash, Qwen3.7-Plus, and the unadapted Qwen3-0.6B model (Base-0.6B). The corresponding PID gains and time-domain metrics are reported in Table 1.

For the delay-dominant FOPDT case C1, the initial controller produces an oscillatory response with an overshoot of 44.58% and an IAE of 437.56. IMC and DeepSeek-V4-Flash reduce the IAE to 214.58 and 237.16, respectively, while keeping the overshoot below 6%. Qwen3.7-Plus nearly eliminates overshoot, reducing it to 0.097%, but produces a slower error-decay pattern and an IAE of 401.63. Base-0.6B largely reproduces the initial gain set and therefore fails to correct the oscillatory response within the eight-round limit.

For the medium-relative-delay FOPDT case C2, the initial overshoot is 54.16%. IMC and both hosted LLMs reduce the IAE from 185.45 to 92.65–102.37 while satisfying the acceptance criteria. Base-0.6B also eliminates the excessive overshoot, reducing it to 0.02%, but its IAE increases to 336.60 because of the substantially slower response. The two FOPDT cases therefore show that multiple controllers can satisfy the same overshoot and steady-state requirements while exhibiting different compromises between damping and error-removal speed.

The SOPDT cases exhibit a different initial-response deficiency. For the delay-dominant C3 process, the initial response has no overshoot but remains too slow to satisfy the steady-state requirement within the simulation horizon. Its steady-state error is 1.11%, and its IAE reaches 625.89. IMC and DeepSeek-V4-Flash satisfy the acceptance criteria in one tuning round, with IAEs of 176.05 and 199.38, respectively. Qwen3.7-Plus requires two rounds and reaches an IAE of 260.15. Base-0.6B produces gains close to the initial settings and provides only limited improvement, although the resulting response satisfies the acceptance criteria.

For the medium-relative-delay SOPDT case C4, all evaluated approaches remove the excessive initial overshoot of 27.57%. IMC gives an overshoot of 7.74% and the lowest IAE of 16.20. DeepSeek-V4-Flash and Base-0.6B produce nearly overshoot-free responses, with overshoots of 1.02% and 0%, respectively, but their IAEs remain close to 30. Qwen3.7-Plus permits a higher but acceptable overshoot of 9.32% and achieves a lower IAE of 23.71. These results again demonstrate that controller acceptance does not imply a unique optimum: different gain combinations can meet the common engineering criteria through different speed–damping trade-offs.

The representative cases also show that the recommended gain adjustments are response dependent rather than governed by a single fixed tuning rule. Depending on the observed deficiency, the models may reduce excessive integral action, modify proportional action, or retain substantial derivative compensation. More importantly, all candidate gains are evaluated using the same response metrics and closed-loop acceptance criteria. Figure 3 therefore demonstrates feasible response correction, rather than establishing that one model or gain pattern is uniformly optimal.

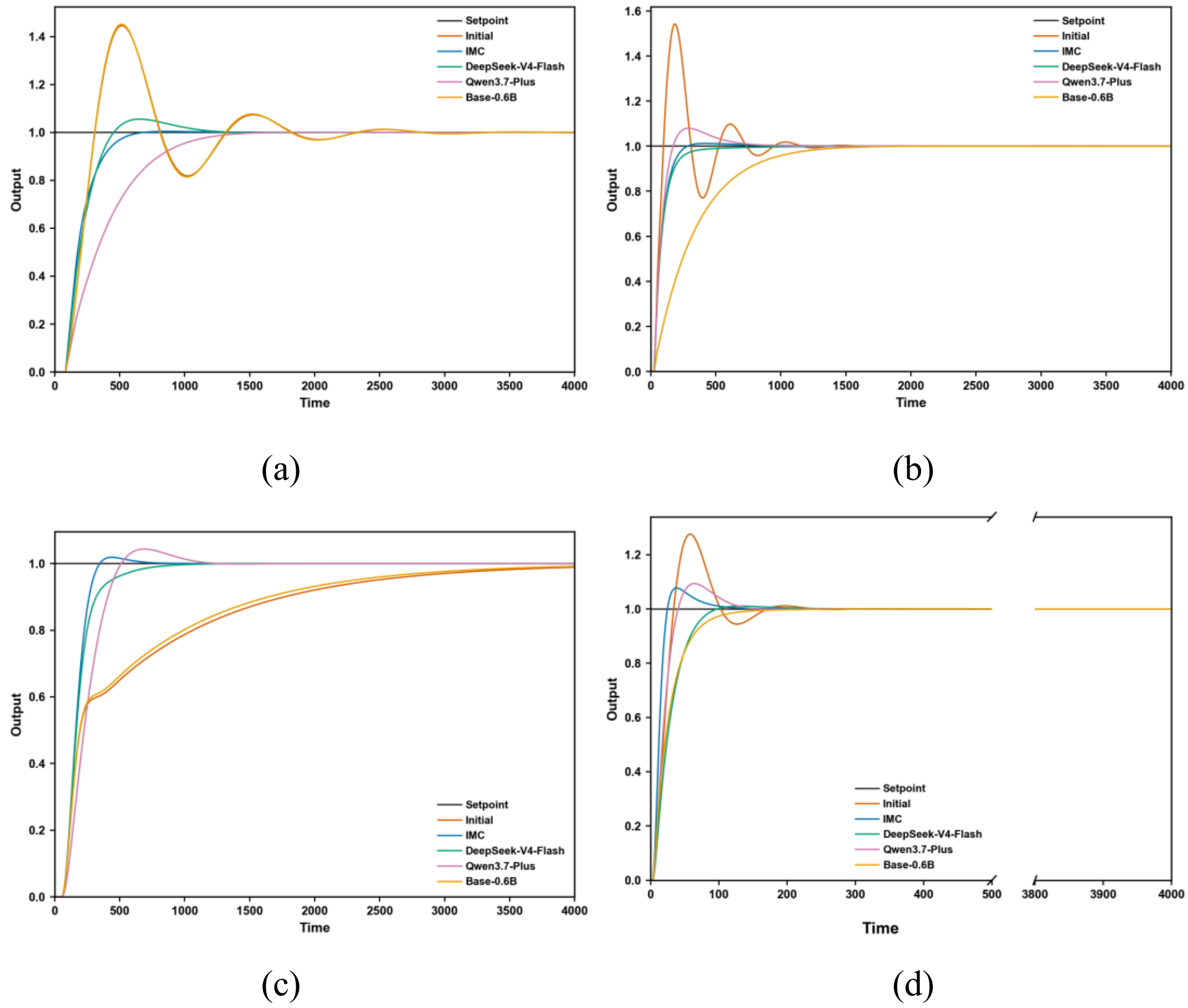

**Figure 3.** Step-response comparison for four representative benchmark cases: (a) $\boldsymbol{G(s)=\frac{0.667}{204.269s+1}e^{-84.060s}}$, (b) $\boldsymbol{G(s)=\frac{0.839}{336.266s+1}e^{-26.350s}}$, (c) $\boldsymbol{G(s)=\frac{0.418e^{-55.630s}}{(54.645s+1)(76.887s+1)}}$, and (d) $\boldsymbol{G(s)=\frac{1.620e^{-2.903s}}{(6.257s+1)(26.231s+1)}}$. Each panel compares the initial controller, IMC, DeepSeek-V4-Flash, Qwen3.7-Plus, and Base-0.6B.

**Table 1.** PID gains and time-domain metrics for the representative benchmark cases.

| Case | Method | $K_p$ | $K_i$ | $K_d$ | IAE | $M_p$ (%) | $e_{ss}$ (%) | Rounds |
|---|---|---|---|---|---|---|---|---|
| C1 | Initial | 0.845 | 0.015 | 61.287 | 437.56 | 44.58 | 0.06 | - |
| C1 | IMC | 1.758 | 0.007 | 61.287 | 214.58 | 0.43 | 0.00 | - |
| C1 | DeepSeek-V4-Flash | 1.500 | 0.008 | 61.287 | 237.16 | 5.56 | 0.00 | 1 |
| C1 | Qwen3.7-Plus | 0.845 | 0.004 | 61.287 | 401.63 | 0.10 | 0.00 | 1 |
| C1 | Qwen3-0.6B-Base# | 0.845 | 0.015 | 61.287 | 441.07 | 45.16 | 0.08 | 8 |
| C2 | Initial | 5.179 | 0.120 | 186.774 | 185.45 | 54.16 | 0.00 | - |
| C2 | IMC | 5.179 | 0.015 | 65.660 | 92.65 | 1.20 | 0.00 | - |
| C2 | DeepSeek-V4-Flash | 5.179 | 0.012 | 60.000 | 99.43 | 0.00 | 0.00 | 1 |
| C2 | Qwen3.7-Plus | 5.179 | 0.024 | 58.730 | 102.37 | 7.93 | 0.00 | 1 |
| C2 | Qwen3-0.6B-Base | 1.451 | 0.003 | 96.702 | 336.60 | 0.02 | 0.00 | 1 |

| Case | Method | $K_p$ | $K_i$ | $K_d$ | IAE | $M_p$ (%) | $e_{ss}$ (%) | Rounds |
|---|---|---|---|---|---|---|---|---|
| C3 | Initial | 1.885 | 0.004 | 60.202 | 625.89 | 0.00 | 1.11 | - |
| C3 | IMC | 1.885 | 0.014 | 60.202 | 176.05 | 1.89 | 0.00 | - |
| C3 | DeepSeek-V4-Flash | 1.885 | 0.012 | 60.202 | 199.38 | 0.00 | 0.00 | 1 |
| C3 | Qwen3.7-Plus | 0.943 | 0.011 | 60.202 | 260.15 | 4.35 | 0.00 | 2 |
| C3 | Qwen3-0.6B-Base# | 1.885 | 0.014 | 60.202 | 590.17 | 0.0 | 0.83 | 1 |
| C4 | Initial | 0.913 | 0.066 | 10.776 | 33.12 | 27.57 | 0.00 | - |
| C4 | IMC | 2.133 | 0.066 | 10.776 | 16.20 | 7.74 | 0.00 | - |
| C4 | DeepSeek-V4-Flash | 0.913 | 0.022 | 10.776 | 29.94 | 1.02 | 0.00 | 1 |
| C4 | Qwen3.7-Plus | 1.250 | 0.045 | 10.776 | 23.71 | 9.32 | 0.00 | 1 |
| C4 | Qwen3-0.6B-Base | 1.359 | 0.021 | 19.207 | 29.70 | 0.00 | 0.00 | 1 |

#Minor IAE differences arise because the initial PID values retain full precision, whereas the reported and model-returned values are rounded to three decimal places.

The batch results in Figure 4 provide a broader evaluation of tuning reliability. Figures 4(a) and 4(d) report the iteration at which the first acceptable controller is obtained for the FOPDT and SOPDT cases, respectively. Figures 4(b) and 4(e) show the final overshoot distributions, while Figures 4(c) and 4(f) show the corresponding steady-state-error distributions.

For the 100 FOPDT cases, DeepSeek-V4-Flash achieves a Pass@8 of 75%, whereas Qwen3.7-Plus reaches 89%. Among successful cases, the mean first-success iterations are 1.37 and 1.21, respectively. Their final median overshoots are 5.88% and 2.10%, and the median steady-state error is essentially zero for both models. Base-0.6B achieves a Pass@8 of only 13%. All 13 successful cases are solved in the first round; none of the remaining 87 cases is recovered through feedback because the model repeats the same PID triplet in every subsequent round. For the 100 SOPDT cases, DeepSeek-V4-Flash and Qwen3.7-Plus achieve Pass@8 values of 79% and 77%, respectively. Their mean first-success iterations are 1.76 and 1.40, while their median overshoots are similar, at 9.24% and 9.29%. Median steady-state error again remains effectively zero. Base-0.6B succeeds in 18% of the cases, and all 82 unsuccessful cases repeat the first-round PID output throughout the eight-round tuning horizon. The similar final success rates of the two hosted models, together with the higher mean iteration count of DeepSeek-V4-Flash, indicate that SOPDT dynamics affect not only whether an acceptable controller can be found but also the correction pathway required to reach it.

A closer examination of the successful Base-0.6B cases further clarifies the origin of its limited tuning capability. After rounding the demonstration gains to the three-decimal precision used in the Base-0.6B outputs, 9 of the 13 successful FOPDT controllers and 16 of the 18 successful SOPDT controllers exactly match one of the PID target triplets provided in the demonstration block. Thus, most of successful cases by Base-0.6B cases can be attributed to direct reuse of a demonstrated PID triplet.

Overall, the representative cases and batch results support the effectiveness of the closed-loop-validated prompting workflow, but they do not establish a universal ranking of the hosted models. Qwen3.7-Plus achieves the higher success rate on the present FOPDT benchmark, whereas DeepSeek-V4-Flash has a small advantage in final SOPDT success. The results also show that an unadapted 0.6B model lacks reliable feedback-conditioned tuning capability, motivating the task-specific adaptation examined in Section 4.2.

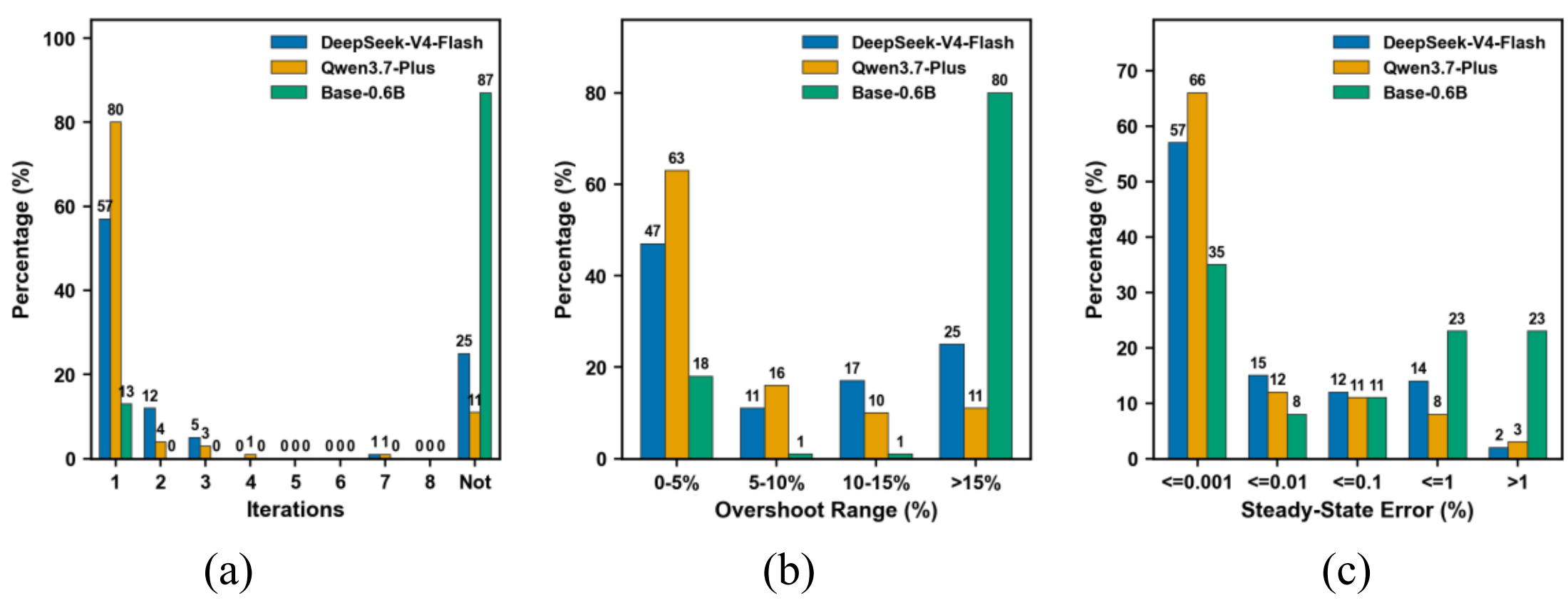

(a) (b) (c)

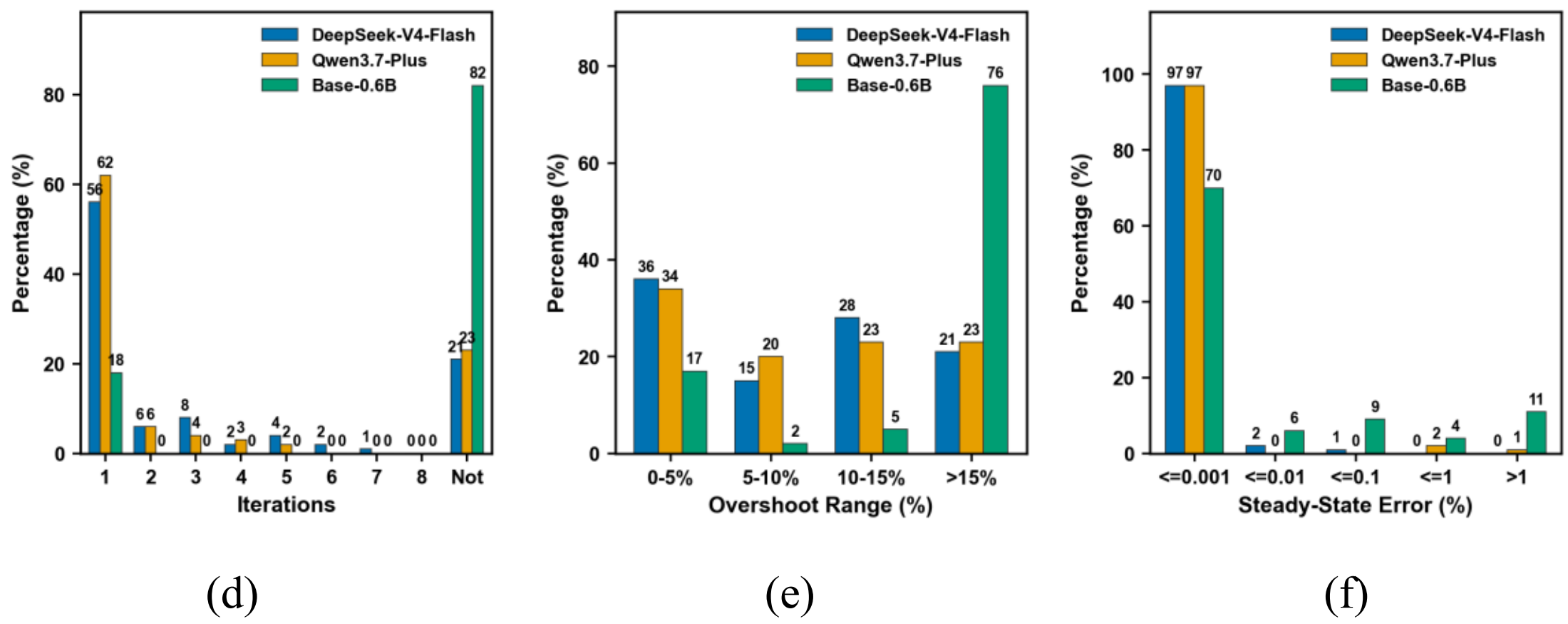

(d) (e) (f)

**Figure 4.** Batch results by DeepSeek-V4-Flash, Qwen3.7-Plus, and Base-0.6B for 100 FOPDT (a-c) and (d-f) SOPDT cases: tuning success, final overshoot, and final steady-state error.

### 4.1.2 Ablation Study on Prompt Strategies

Prompt ablation is evaluated on the same 200 benchmark cases using three inputs: the complete prompt (Full), three headline indicators only (KPI-3), and all eight numerical indicators without diagnostic descriptions (Numeric-8). Table 2 summarizes the success rate. For the hosted models, semantic descriptions are not uniformly beneficial. On FOPDT processes, DeepSeek-V4-Flash reaches 75% success with Full, compared with 89% for KPI-3 and 86% for Numeric-8; Qwen3.7-Plus reaches 89%, 81%, and 89%, respectively. The simpler first-order dynamics can evidently be inferred from the numerical indicators by a capable model. On SOPDT processes, however, removing information is less reliable: DeepSeek-V4-Flash decreases from 79% with Full to 62% with Numeric-8, while Qwen3.7-Plus decreases to 68% with KPI-3, although its Numeric-8 result (79%) remains close to Full (77%).

The capacity dependence becomes clearer for the unadapted Qwen3-0.6B model. All three prompts exhibit a floor effect on FOPDT processes (13 – 17% Pass@1), and later feedback rounds do not add successes because unsuccessful cases repeat the first PID output. On SOPDT processes, Full raises Pass@1 to 18%, compared with 5% for KPI-3 and 7% for Numeric-8.

This improvement should not be interpreted as a safety guarantee. For Base-0.6B, the complete prompt also produces more aggressive but incorrect gain changes in some cases, including severe IAE regressions. Semantic guidance can improve response interpretation, but it cannot determine whether the resulting controller is stable or effective.

Full is therefore retained as the unified prompt for subsequent experiments. It preserves the complete measured response, provides the strongest support for the capacity-limited SOPDT setting, and maintains a common interface across model families. The ablation study does not show that Full maximizes every metric for every process. Rather, it shows that semantic structure becomes most valuable when either model capacity or process complexity limits inference from numerical indicators alone.

**Table 2.** Prompt-ablation success rates for the 200-case benchmark. Hosted-model results are reported as Pass@8; Base-0.6B results are reported as Pass@1 because subsequent feedback rounds do not produce additional successful cases. Detailed statistics can be found in Supporting Information S7.

| Process | Model | Full | KPI-3 | Numeric-8 |
|---|---|---|---|---|
| FOPDT | DeepSeek-V4-Flash | 75 | 89 | 86 |
| FOPDT | Qwen3.7-Plus | 89 | 81 | 89 |
| FOPDT | Base-0.6B | 13 | 16 | 17 |
| SOPDT | DeepSeek-V4-Flash | 79 | 78 | 62 |
| SOPDT | Qwen3.7-Plus | 77 | 68 | 79 |
| SOPDT | Base-0.6B | 18 | 5 | 7 |

#### 4.1.3 Dynamic Prompting

Dynamic demonstration libraries are used to encode tuning preferences without modifying the model weights. DeepSeek-V4-Flash is evaluated on the two delay-dominant cases C1 and C3 using conservative, balanced, and aggressive IMC-based demonstrations. Larger IMC filter multipliers favor smoother and more robust responses, whereas smaller multipliers favor faster and more aggressive tracking. Figure 5 compares the resulting responses, and Table 3 summarizes the corresponding PID gains and performance metrics.

Changing the demonstration library produces systematic shifts in both controller gains and closed-loop behavior. For C1, conservative and balanced prompting yield acceptable controllers, with the balanced strategy providing the best overall compromise between IAE and overshoot. Aggressive prompting increases the gain scale and accelerates the response, but the overshoot rises to 45.78%. A similar trade-off is observed for C3: conservative and balanced demonstrations produce overshoot-free responses, whereas aggressive prompting gives the lowest IAE but increases overshoot to 16.53%. Both aggressive controllers are therefore rejected under the common 15% overshoot criterion.

These results show that dynamic demonstrations can steer the model toward different speed–robustness preferences. However, they act as soft conditioning signals rather than guarantees of an admissible response. The demonstration library specifies the desired tuning tendency, while closed-loop validation determines whether the generated controller satisfies the engineering acceptance criteria.

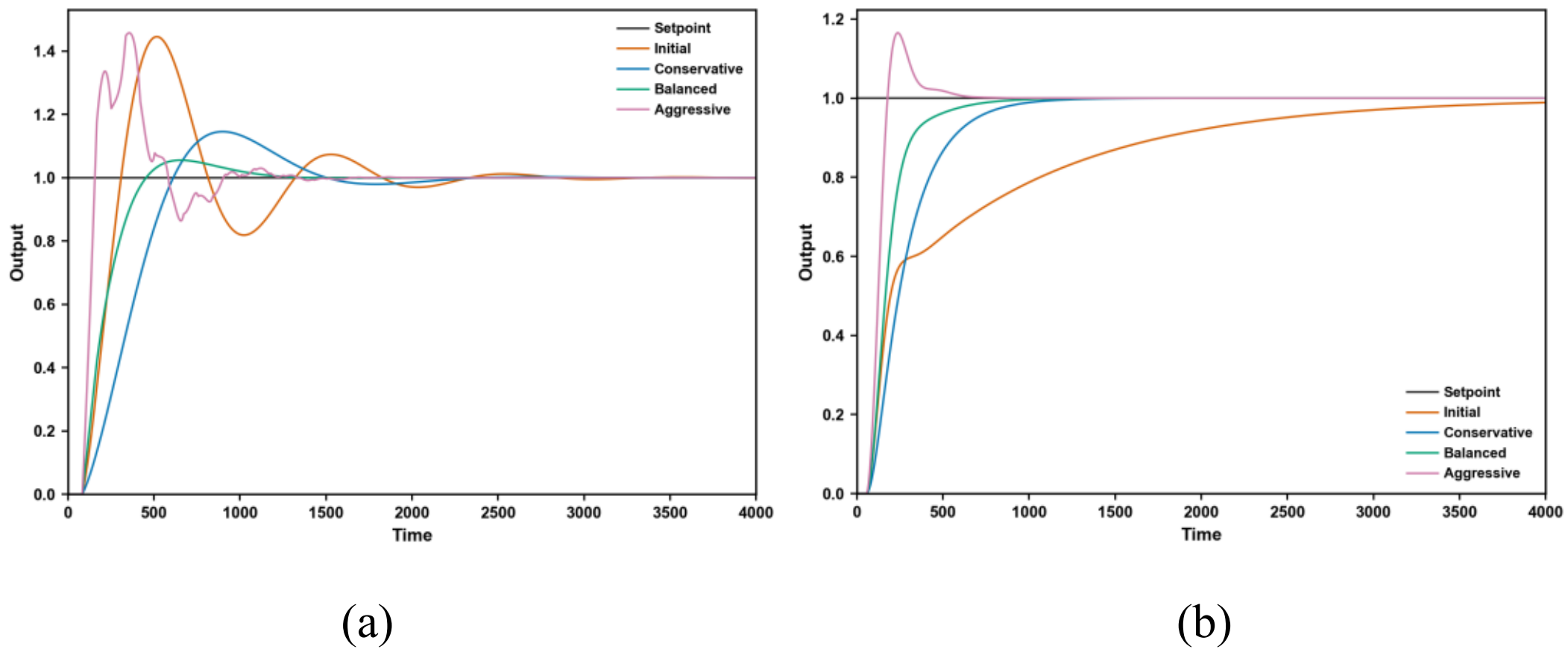

(a) (b)

**Figure 5.** Effect of conservative, balanced, and aggressive demonstration libraries on DeepSeek-V4-Flash tuning for the delay-dominant representative cases: (a) FOPDT case C1 and (b) SOPDT case C3.

**Table 3.** PID gains and closed-loop performance metrics obtained using conservative, balanced, and aggressive demonstration libraries for cases C1 and C3.

| Case | Preference | $K_p$ | $K_i$ | $K_d$ | IAE | $M_p$ (%) | $e_{ss}$ (%) | Rounds |
|---|---|---|---|---|---|---|---|---|
| C1 | Conservative | 0.350 | 0.006 | 18.000 | 426.32 | 14.58 | 0.01 | 4 |
| C1 | Balanced | 1.500 | 0.008 | 61.287 | 237.16 | 5.56 | 0.00 | 1 |
| C1 | Aggressive | 3.600 | 0.036 | 250.000 | 250.23 | 45.78 | 0.00 | 8 |
| C3 | Conservative | 0.943 | 0.008 | 30.101 | 299.02 | 0.00 | 0.00 | 1 |
| C3 | Balanced | 1.885 | 0.012 | 60.202 | 199.38 | 0.00 | 0.00 | 1 |
| C3 | Aggressive | 3.574 | 0.024 | 131.580 | 148.17 | 16.53 | 0.00 | 8 |

## 4.2 Performance of Fine-Tuned Small Language Models

### 4.2.1 SFT and GRPO Training

Figure 6(a) shows the training trajectory during full-parameter SFT. Both the training and validation losses decrease rapidly during the early stage and subsequently approach a common plateau. The validation loss decreases from 0.767 at step 50 to 0.639 after the first epoch and stabilizes at approximately 0.623, while the training loss follows a similar trend and reaches 0.619 at the end of training. The small and stable gap between the two curves, together with the absence of a sustained increase in validation loss, indicates stable supervised learning without evident overfitting. Most of the improvement occurs during the first epoch, with later epochs providing gradual refinement.

Figures 6(b)–(d) summarize the PI-GRPO training dynamics. Because process instances and candidate controllers are sampled online, the instantaneous reward varies substantially across optimization steps. Nevertheless, its moving average increases rapidly during the early stage and reaches its highest level near step 100. The KL divergence follows a similar initial increase and is subsequently moderated toward a narrower range by the adaptive KL mechanism. Evaluation on a fixed validation set provides a clearer basis for checkpoint selection. At step 100, Pass@1 and the mean IAE-improvement reward reach their joint maxima of 54.5% and 0.330, respectively. Continued optimization increases neither metric: by step 800, they decrease to 19.0% and 0.057. Correspondingly, validation success declines from 109 of 200 cases at step 100 to 38 of 200 cases at step 800, indicating over-optimization toward the online training distribution. The step-100 checkpoint is therefore selected for all subsequent evaluations rather than the final training state.

By step 100, the five-GPU training run has evaluated approximately 2,000 online prompts and 8,000 candidate completions. Starting from the SFT initialization, these interactions are sufficient to produce a measurable improvement in first-recommendation performance. More importantly, the divergence between online reward and fixed-set validation demonstrates that online reward alone is not a reliable criterion for selecting the final PI-GRPO checkpoint.

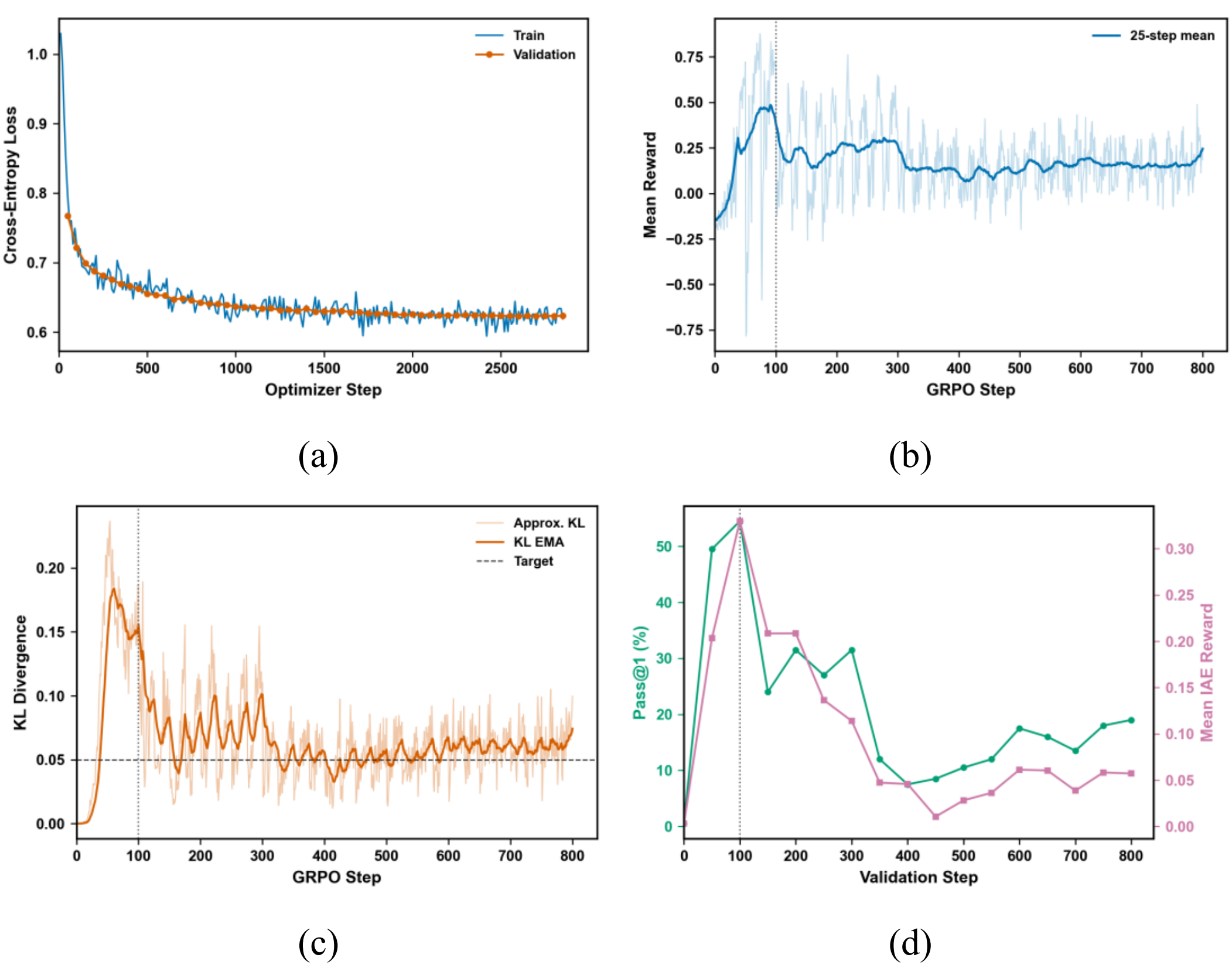

**Figure 6.** Training dynamics of Qwen3-0.6B. (a) Training and validation cross-entropy losses during full-parameter SFT. PI-GRPO training dynamics: (b) instantaneous online reward and its 25-step moving average; (c) approximate KL divergence, its exponential moving average, and the prescribed KL target; and (d) Pass@1 and mean IAE-improvement reward on the fixed 200-process validation set. The vertical dotted line indicates the selected checkpoint at step 100. Validation in panel (d) uses greedy decoding with the SFT backbone loaded in 4-bit

NF4 precision; the reported values therefore characterize checkpoint evolution within the quantized training environment rather than the absolute performance of the final deployment configuration.

#### 4.2.2 Results of SFT and GRPO models

**Figure 7** compares aggregate Pass@1 across all 200 benchmark cases, thereby isolating the quality of the first PID recommendation. This metric is particularly relevant to low-latency deployment, where an acceptable controller is expected without repeated model calls. Base-0.6B achieves a Pass@1 of only 15.5%. SFT increases this value to 86.5%, and PI-GRPO further improves it to 94.0%. Relative to SFT, PI-GRPO produces 20 newly successful cases while losing success on five cases. The advantage of PI-GRPO narrows at longer horizons: Pass@2 is 94% versus 95%, and both models reach a final success rate of 95%. PI-GRPO therefore improves the probability of obtaining an acceptable controller on the first recommendation rather than substantially expanding the final set of solvable cases. Both adapted 0.6B models also outperform the hosted LLM baselines in Pass@1: DeepSeek-V4-Flash and Qwen3.7-Plus achieve 56.5% and 71.0%, respectively. These results show that task-specific supervision and physics-informed policy optimization can compensate for the performance gap associated with the smaller model scale. Step-response comparison for four representative benchmark cases can be found in Supporting Information S9.

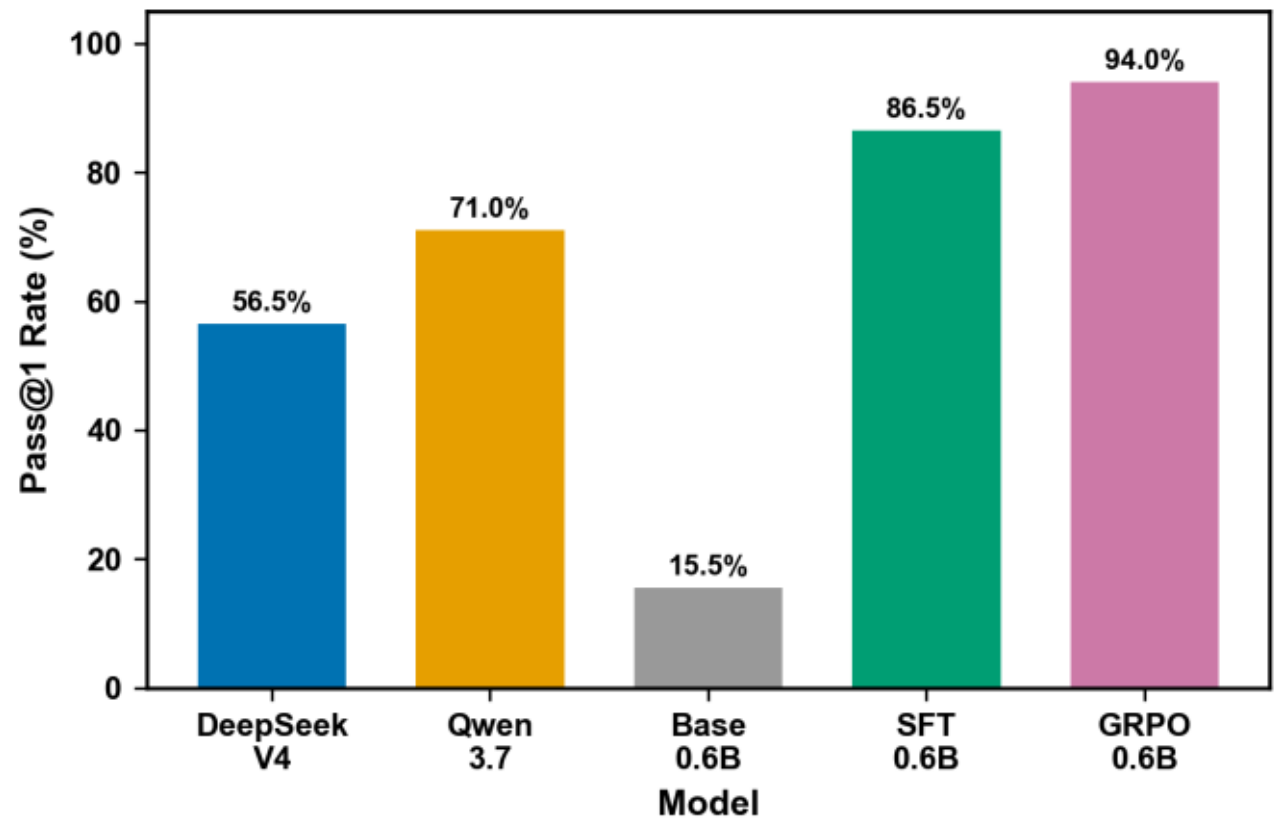

**Figure 7.** One-step benchmark across DeepSeek-V4-Flash, Qwen3.7-Plus, Base-0.6B, SFT-0.6B, and GRPO-0.6B for FOPDT and SOPDT cases. Overshoot and steady-state error can be found in Supporting Information S8.

### 4.2.3 Robustness–Performance Trade-Off

The exact-delay frequency-domain analysis in Figures 8 and 9 compares the initial, SFT-0.6B, and GRPO-0.6B controllers and clarifies how model adaptation changes the balance among response speed, damping, and robustness.

For the selected FOPDT case C1, the initial controller has a phase margin of 32.94° at a gain crossover frequency of 0.00585 rads$^{-1}$. This limited phase margin is consistent with the lightly damped response and the large overshoot of 44.58%. SFT increases the phase margin to 76.21°, reduces the crossover frequency to 0.00481 rads$^{-1}$, and lowers the IAE from 437.56 to 217.27. PI-GRPO further increases the phase margin to 80.48° and reduces the overshoot from 1.44% to 0.16%. However, relative to SFT, the crossover frequency decreases by approximately 9.8%, while the IAE increases by 8.8%, from 217.27 to 236.33. The Nyquist trajectories remain separated from the critical point (−1,0), consistent with the positive phase margins and stable time-domain responses. Thus, PI-GRPO does not uniformly outperform SFT, but exchanges a modest reduction in response speed for greater damping and robustness.

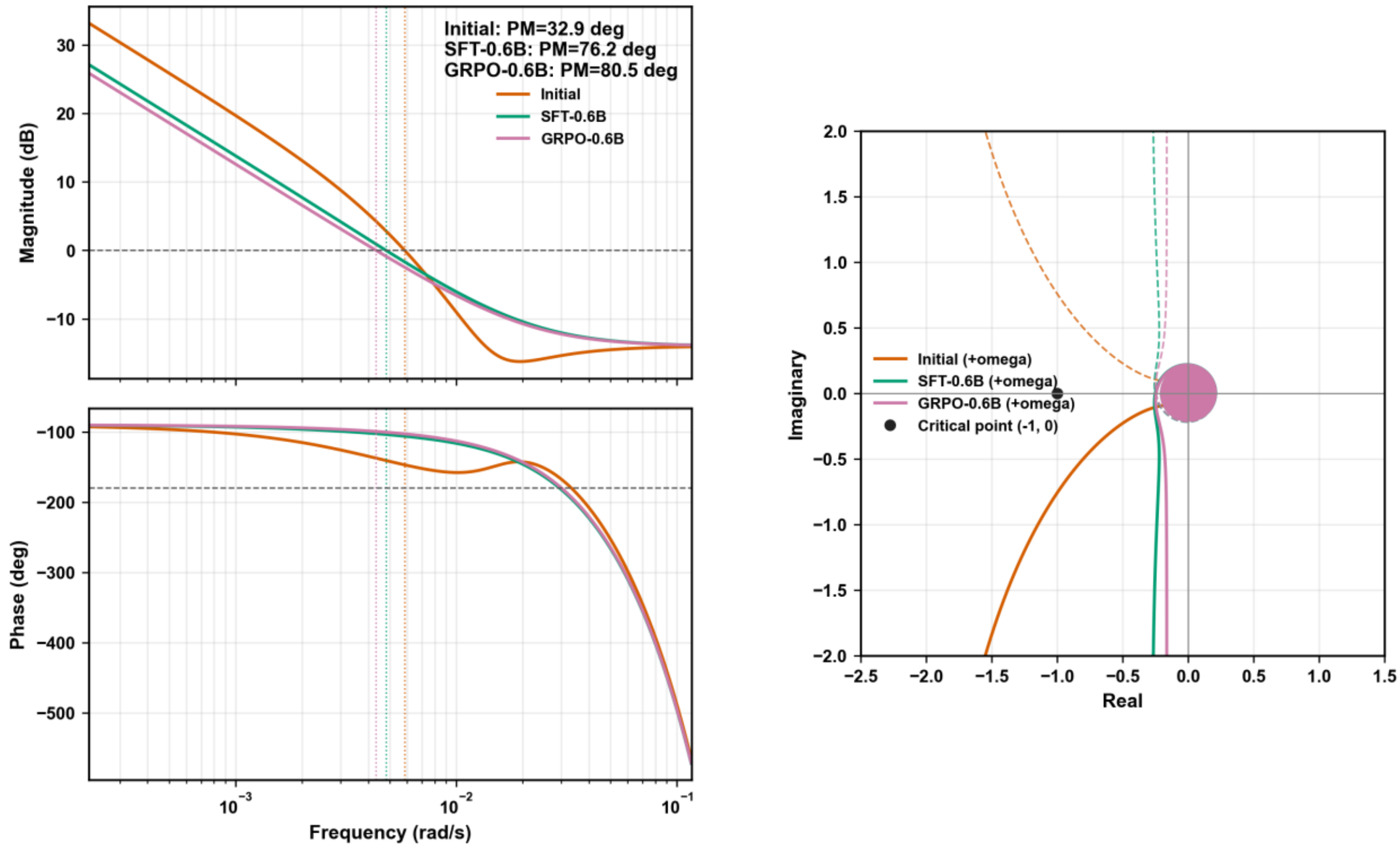

**Figure 8.** Exact-delay Bode and Nyquist comparison of the initial, SFT-0.6B, and GRPO-0.6B controllers for the selected FOPDT case C1.

The selected SOPDT case C3 exhibits a different progression. The initial controller has the largest phase margin, 116.76°, but a low crossover frequency of only 0.00222 rads$^{-1}$. Its large stability margin therefore reflects an overly conservative controller rather than superior overall performance, as indicated by an IAE of 625.89 and a steady-state error of 1.11% over the evaluation horizon. SFT increases the crossover frequency to 0.00666 rads$^{-1}$ and reduces the IAE to 185.05, but the phase margin decreases to 62.87°, and the response exhibits 9.67% overshoot. PI-GRPO moderates this trade-off by increasing the phase margin to 78.06° and eliminating overshoot, while reducing the crossover frequency to 0.00540 rads$^{-1}$. This improvement in damping is accompanied by a 6.0% increase in IAE relative to SFT, from 185.05 to 196.11.

Together, the two cases show that SFT primarily improves task performance, while PI-GRPO shifts the adapted policy toward higher stability margins and lower overshoot, with a moderate cost in bandwidth and accumulated error. Because the frequency-domain calculations retain the exact delay term, they provide a robustness characterization complementary to the Padé-based Routh–Hurwitz screening used during training. In both cases, PI-GRPO refines the robustness–performance balance of an already stable SFT controller rather than converting an unstable controller into a stable one.

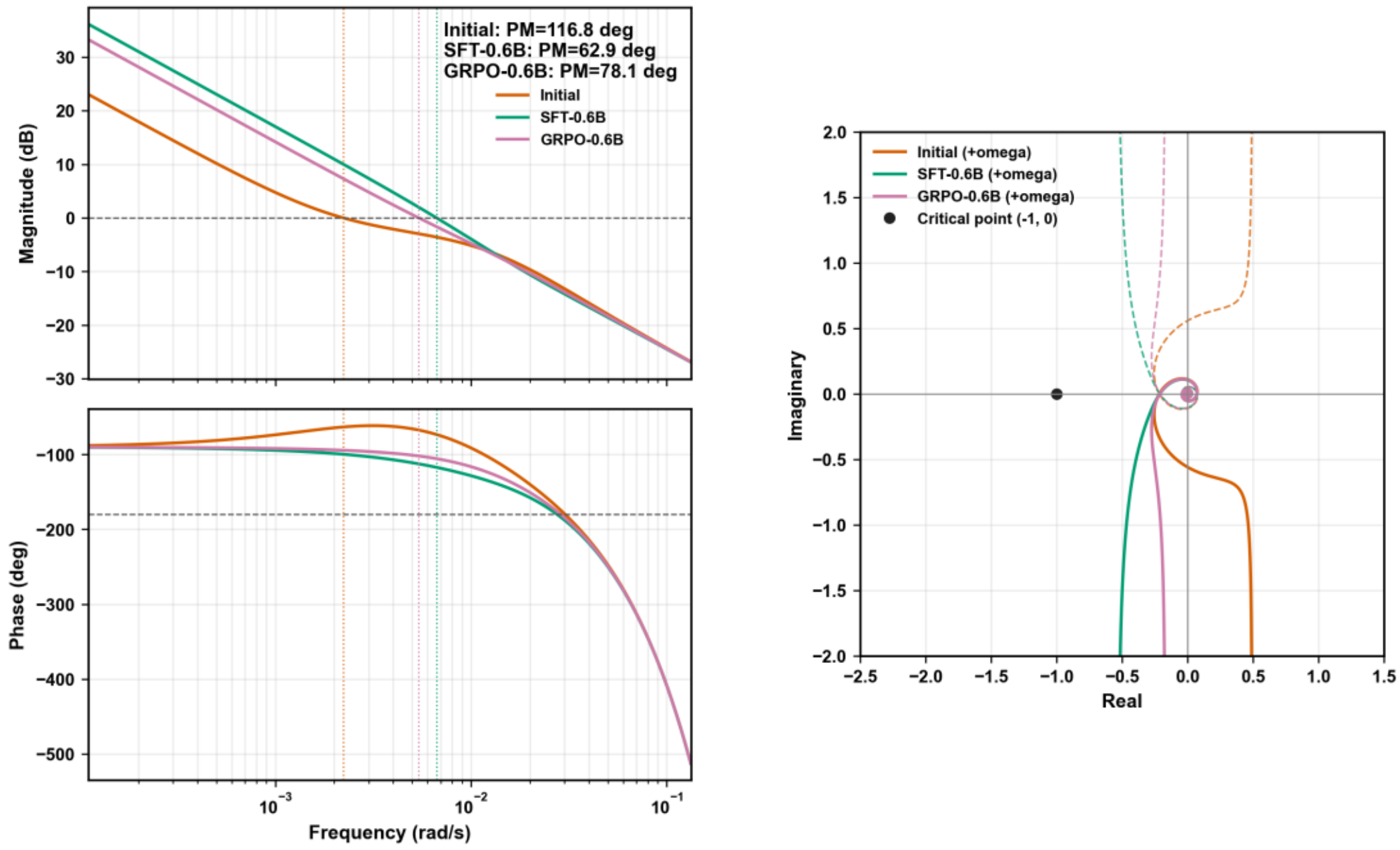

**Figure 9.** Exact-delay Bode and Nyquist comparison of the initial, SFT-0.6B, and GRPO-0.6B controllers for the selected SOPDT C3.

### 4.3 Deployment Implications

The results support deployment feasibility at the model level. After task adaptation, the 0.6B model achieves high first-recommendation success and can be executed locally on a single GPU. However, the present benchmark does not establish plant readiness. Actuator limits, measurement noise, model mismatch, process uncertainty, failure cases, and real-time latency must still be evaluated. Industrial deployment should therefore retain an analytical and closed-loop validation layer rather than apply language-model-generated gains without verification.

## 5. Conclusion

This study develops a closed-loop-validated language-model framework for PID tuning of delayed processes. On 100 FOPDT and 100 SOPDT cases, the hosted LLMs achieve final success rates of 75–89% and 77–79%, respectively. Prompt ablation shows that semantic diagnostics are not universally superior but become valuable when model capacity or process complexity is limiting. Dynamic demonstrations further enable controllable speed–robustness preferences.

Building on the same response representation and acceptance criteria, the framework is further extended to lightweight deployment through full-parameter SFT and PI-GRPO. SFT increased the Pass@1 of Qwen3-0.6B to 86.5%, and PI-GRPO further raised it to 94.0%, while both adapted models achieved a Pass@8 of 95%. PI-GRPO mainly improved first-recommendation reliability and stability margins, although these gains are sometimes accompanied by reduced bandwidth and higher IAE. Together, these results demonstrate that the proposed framework can support both hosted-model reasoning and lightweight local tuning under a common validation logic. Future work will extend the approach to nonlinear and multivariable plants and evaluate actuator constraints, measurement noise, model uncertainty, and real-time deployment performance.

**Data Availability Statement**

The code supporting the findings of this study are publicly available at https://github.com/PSEZJUCE/LLMPIDTuner-public. The datasets, evaluation results, training records, and frozen experimental protocol supporting this study are available in the Zenodo repository at https://doi.org/10.5281/zenodo.21669697.

**Acknowledgments**

The financial support provided by the Project of the National Natural Science Foundation of China (22308314, U22A20415), and the Natural Science Foundation of Zhejiang Province, China (LQ24B060001) is gratefully acknowledged. The authors also thank the AI + High Performance Computing Center of ZJU-ICI.

**Author Contributions**

Zhoupeng Shou: Conceptualization, Methodology, Software, Validation, Data curation, Writing - original draft. Xiaodong Hong: Conceptualization, Supervision, Writing - review & editing. Congjing Ren: Methodology, Validation, Writing - review & editing. Jingdai Wang: Resources, Supervision. Yongrong Yang: Supervision, Project administration. Zuwei Liao: Funding acquisition, Supervision, Project administration, Writing - review & editing. All authors have read and approved the final manuscript.

**Supporting Information**

**S1.** Performance Indicators and Prompt Representation. **S2.** Process Simulation Environment. **S3.** IMC Demonstration Cases. **S4.** SFT and GRPO Settings. **S5.** Physics-Informed Reward Function. **S6.** Evaluation Protocol and Transfer-Function Pool. **S7.** Ablation Study on Prompt Strategies. **S8.** Detailed Statistics by SFT-0.6B and GRPO-0.6B. **S9.** Response Curves by SFT-0.6B and GRPO-0.6B. S10. FOPDT and SOPDT Benchmarks.

# Supporting Information

**S1. Performance Indicators and Prompt Representation**

This section defines the response quantities shared by prompt construction, benchmark evaluation, SFT-data generation, and PI-GRPO reward calculation. Let $r$ be the setpoint, $y_t$ the sampled process output, $e_t = r - y_t$ the tracking error, $\Delta t$ the sampling interval, and $N$ the number of samples.

The full prompt representation contains IAE and seven response features: overshoot, steady-state error, settling time, the number of significant setpoint crossings, attenuation ratio, oscillation period, time to 63.2% of the setpoint after the dead time. IAE is evaluated with the same rectangular sum used by the simulator:

$$IAE = \sum_{k=0}^{N-1} |e_t| \times \Delta t \tag{S1}$$

Overshoot is computed from the largest sampled output defined in the manuscript. Steady-state error is the relative error between the setpoint and the mean of the final $m = \min(100, N)$ samples, defined in the manuscript. Settling time $T_s$ is the first sample after the last excursion outside a ±5% band; if the final sample lies outside this band, the response is classified as unsettled.

A scale factor $s_r = \max(|r|, 10^{-12})$ is used for numerical protection. The number of significant setpoint crossings is calculated after discarding samples within the $\pm 5\backslash\%$ band, i.e., $|y_k - r| \le 0.05 s_r$. For the remaining ordered deviations $(\tilde{e_j})$, the crossing count is

$$N_{\text{cross}} = \sum_{j=2}^{m} [\text{ign}(\tilde{e_j}) \neq \text{sign}(\widetilde{e_{j-1}})] \tag{S2}$$

And $N_{\text{cross}}$ is set to zero when fewer than two significant samples remain.

Local output maxima are detected over the complete simulated response using a minimum prominence of $0.01 s_r$. If the first two detected peaks occur at indices $p_1$ and $p_2$, the attenuation ratio $A_r$ and oscillation period $T_{\text{osc}}$ are calculated as

$$A_r = \frac{|y_{p_2} - r|}{\max(|y_{p_1} - r|, 10^{-12})}, T_{\text{osc}} = t_{p_2} - t_{p_1} \tag{S3}$$

Both quantities are set to zero when fewer than two peaks are detected. Finally, because all experiments use $r = 1$, the time to 63.2% after the dead time is

$$t_{63.2} = \max(0, t_{k^*} - \theta), k^* = \min\{k : t_k \ge \theta,\ y_k \ge 0.632 r\} \tag{S4}$$

No interpolation is applied between adjacent samples. If the threshold is not reached within the simulation horizon, $t_{63.2}$ is reported as unavailable.

The numerical indicators are converted into deterministic engineering descriptions. The rules distinguish excessive overshoot, repeated crossings, weak attenuation, slow response, and persistent error. Numerical values are retained alongside the descriptions so that all model families receive the same quantitative evidence and the same diagnostic interpretation. An initial prompt contains a frozen demonstration block and a current-state block. The current state reports the PID gains, IAE, observed dead time, requested control style, and response diagnosis. A feedback prompt reports the response obtained after applying the previous recommendation and requests another gain set. The canonical output is P:<Kp>; I:<Ki>; D:<Kd>.

**S2. Process Simulation Environment**

The shared simulator evaluates an ideal parallel PID controller on FOPDT and SOPDT processes with a pure dead time. It is used without changing numerical definitions across LLM evaluation, SFT-data construction, PI-GRPO reward calculation, and final model comparison.

$$G(s) = \frac{K \cdot e^{-\theta s}}{Ts+1} \tag{S5}$$

where $K$ is the process gain, $T$ is the FOPDT time constant, and $\theta$ is the process dead time. The SOPDT systems are implemented using an equivalent two-time-constant form:

$$G(s) = \frac{K \cdot e^{-\theta s}}{(\tau_1 s+1)(\tau_2 s+1)} \tag{S6}$$

where $\tau_1$ and $\tau_2$ are the two lag time constants.

The controller is implemented in parallel position form. The integral state is updated by rectangular accumulation and the derivative term uses the current error difference. At the initial sample, the derivative contribution is set to zero to avoid an artificial derivative kick.

$$u_k = K_p e_k + K_i I_k + K_d\left(e_k - e_{(k-1)}\right)/\Delta t, I_k = I_{(k-1)} + e_k \Delta t \tag{S7}$$

The process dynamics are integrated using fixed-step, first-order Euler updates with $\Delta t = 0.1\ s$. A non-integer dead time is represented by linear interpolation of the stored controller-output history. For an SOPDT process, a sequential Euler discretization is used: the first-lag state is advanced using the delayed controller output, after which the second-lag state is advanced using the newly updated first-lag state. This update order is held fixed across all simulations.

Specifically, the SOPDT states are updated as

$$x_{1,n+1} = x_{1,n} + \frac{\Delta t}{\tau_1}\left(K u_{d,n} - x_{1,n}\right), x_{2,n+1} = x_{2,n} + \frac{\Delta t}{\tau_2}\left(x_{1,n+1} - x_{2,n}\right), y_{n+1} = x_{2,n+1} \tag{S8}$$

where $u_{d,n}$ denotes the delayed controller output evaluated at $t_n - \theta$. When $\theta/\Delta t$ is non-integer, $u_{d,n}$ is obtained by linear interpolation of the stored controller-output history; $u_{d,n} = 0$ before the delayed input becomes available. Thus, the second lag uses the first-lag state updated within the same time step rather than its value from the preceding step.

The setpoint is 1, the simulation horizon is 4000 s, and 40,001 points include both endpoints. The process-specific dead time replaces the default simulation value. If the output, error, or control signal becomes non-finite, or if $|y|$ exceeds 3, the trajectory is marked non-finite and is not accepted as converged. Initial PID gains are generated by the protocol described in Section S6 rather than fixed to a common triplet.

**S3. IMC Demonstration Cases**

Demonstration cases provide concrete examples of response diagnosis and gain correction. Independent sets of ten FOPDT and ten SOPDT source cases are generated before construction of the benchmark sets and are then held fixed. Each set spans the ten controller-defect types and the four dead-time strata used by the experimental protocol.

For each source case, an IMC reference controller is calculated using $\lambda = m \max(\theta, 0.1T_c)$, where $T_c = T$ for FOPDT and $T_c = \tau_1 + \tau_2$ for SOPDT. The multipliers $m$ are 0.8, 2.0, and 5.0 for aggressive, balanced, and conservative styles, respectively.

$$\lambda_{aggressive} = 0.8 \max(\theta, 0.1T_c), \lambda_{balanced} = 2 \max(\theta, 0.1T_c), \lambda_{conservative} = 5.0 \max(\theta, 0.1T_c) \quad \text{(S9)}$$

The balanced/full version is used in the principal comparisons, SFT, and PI-GRPO. The same source cases are rendered into five prompt assets: balanced/full, balanced/kpi3, balanced/numeric8, aggressive/full, and conservative/full. The three balanced variants differ only in the amount and form of response information. The style variants retain the same initial cases but replace the target controller with the corresponding IMC style.

**S4. SFT and GRPO Settings**

SFT uses 40,000 chat-format records following the OpenAI message schema: 20,000 FOPDT and 20,000 SOPDT records generated with seed 81001. All records use the balanced/full demonstration and a balanced, simulation-verified IMC target. The prompt does not reveal the transfer-function parameters; it contains the current PID gains, IAE, observed dead time, requested style, and response diagnosis. Approximately half of the records include one to three intermediate feedback turns before the final target.

Only assistant tokens contribute to the autoregressive SFT loss. Full-parameter training uses bf16 precision, a maximum sequence length of 6144, a learning rate of $1 \text{x} 10^{-5}$, three epochs, one sequence per

GPU, and eight-step gradient accumulation. The AdamW weight decay is 0.01, gradients are clipped at 1.0, and the learning-rate warmup ratio is 0.03. Five percent of the records form a validation split stratified by process type and sample kind; evaluation and checkpointing occur every 50 optimizer steps, and the lowest-loss checkpoint is restored. Records exceeding the token limit are rejected during preflight validation rather than silently truncated.

After SFT, physics-informed group relative policy optimization (GRPO) is applied to improve the model beyond imitation of the IMC-derived targets. For each online prompt $q$, the rollout policy $\pi_{\Theta_{\text{old}}}$ samples a group of $G$ candidate completions $\{o_i\}_{i=1}^{G}$, each containing one complete PID parameter set. In this study, $G$ is set to 4. Each candidate is evaluated in the same closed-loop environment and assigned a scalar physics-informed reward $R_i$. The token-level GRPO objective is defined as follows.

$$\mathcal{J}_{\text{GRPO}}(\Theta) = E_{q\sim\mathbb{D}}\left[E_{\{o_i\}_{i=1}^{G}\sim\pi_{\Theta_{\text{old}}}(\cdot|q)}\left[\mathcal{J}\left(q,\{o_i\}_{i=1}^{G}\right)\right]\right] \tag{S10a}$$

$$\mathcal{J}\left(q,\{o_i\}_{i=1}^{G}\right) = \left[\frac{1}{\sum_{i=1}^{G} Token_i}\sum_{i=1}^{G}\sum_{t=1}^{Token_i}\left(\min\left(\rho_{i,t}(\Theta)A_i, \text{clip}\left(\rho_{i,t}(\Theta), 1-\epsilon, 1+\epsilon\right)A_i\right) - \beta_k\widehat{D_{\text{KL},i,t}}(\Theta)\right)\right] \tag{S10b}$$

$$\rho_{i,t}(\Theta) = \frac{\pi_\Theta(o_{i,t}|q, o_{i,<t})}{\pi_{\Theta_{\text{old}}}(o_{i,t}|q, o_{i,<t})} \tag{S10c}$$

$$\widehat{D_{\text{KL},i,t}}(\Theta) = \exp\left(l_{i,t}^{\text{ref}} - l_{i,t}^{\Theta}\right) - \left(l_{i,t}^{\text{ref}} - l_{i,t}^{\Theta}\right) - 1 \tag{S10d}$$

$$l_{i,t}^{\Theta} = \log\pi_\Theta\left(o_{i,t} \mid q, o_{i,<t}\right), l_{i,t}^{\text{ref}} = \log\pi_{\text{ref}}\left(o_{i,t} \mid q, o_{i,<t}\right) \tag{S10e}$$

Here, $q$ denotes an online prompt sampled from the prompt distribution $\mathbb{D}$. The completion $o_i = \left(o_{i,1}, \dots, o_{i,Token_i}\right)$ is the $i$-th candidate generated for $q$, and $Token_i$ is its number of valid completion tokens. The outer expectation averages over sampled prompts, whereas the inner expectation accounts for the stochastic generation of candidate completions conditioned on the same prompt. The candidate-level advantage $A_i$ is shared by all valid tokens in completion $o_i$.

The quantity $\rho_{i,t}(\Theta)$ is the probability ratio for the $t$-th token of candidate $i$ between the current policy and the rollout policy. The clipping threshold is $\epsilon = 0.2$. The adaptive coefficient $\beta_k$ is fixed for all candidates and tokens within optimization step $k$ and is updated between optimization steps. The term $\widehat{D_{\text{KL},i,t}}(\Theta)$ is the token-level KL estimator used to penalize deviation from the frozen SFT reference policy $\pi_{\text{ref}}$. $\text{clip}(x, l, u)$ truncates $x$ to the interval $[l, u]$. The group-relative advantage $A_i$ is obtained by normalizing candidate rewards within the same prompt group:

$$A_i = \frac{R_i - \bar{R}}{\sigma_R},\ \ \bar{R} = \frac{1}{G}\sum_{i=1}^{G} R_i,\ \ \sigma_R = \sqrt{\frac{1}{G}\sum_{i=1}^{G}(R_i - \bar{R})^2} \tag{S11}$$

The population standard deviation is used for group normalization. When $\sigma_R \leq 10^{-6}$, all advantages in that candidate group are set to zero.

PI-GRPO generates balanced first-turn prompts online rather than reading a fixed prompt dataset. At each optimization step, each of the five distributed ranks samples four online prompts, and the rollout policy generates four candidate completions per prompt. An uninterrupted 800-step run therefore processes 16,000 prompt instances and evaluates 64,000 candidate completions. Therefore, an uninterrupted 800-step run processes $(800 \times 5 \times 4 = 16{,}000)$ prompt instances and evaluates $(16{,}000 \times 4 = 64{,}000)$ candidate completions. Sampling uses temperature = 0.9, top-p = 0.95, and top-k = 50.

The policy is updated with QLoRA using 4-bit NF4 quantization, $r = 16$, $\alpha = 32$, zero LoRA dropout, and bf16 computation. AdamW uses zero weight decay, a maximum gradient norm of 1.0, and micro-batches of four completions. The learning rate increases linearly to $1\times10^{-5}$ during the first 50 steps and then follows cosine decay to $1\times10^{-6}$ at step 800. Prompt and completion limits are 6144 and 64 tokens.

The KL exponential moving average uses coefficient 0.10. Every ten steps, beta is multiplied by 1.5 when the KL exponential moving average exceeds 0.10 and divided by 1.5 when it falls below 0.025, subject to the stated bounds. Exceeding 0.20 for 25 consecutive steps triggers an emergency checkpoint and stop. Validation checkpoints are compared lexicographically by Pass@1, mean bounded IAE-improvement score, mean reward, and lower KL; if none surpasses step 0, the SFT model remains selected. Checkpoints preserve model, optimizer, framework random state, adaptive-KL state, validation history, and each rank's online-prompt random state. Exact continuation requires the same number of ranks and an otherwise unchanged training configuration.

**Table S1 Key hyperparameter settings for two-stage training**

| Stages | Parameters | Setting/Value |
|---|---|---|
| SFT | Fine-tuning method | Full |
| SFT | Training data | 40,000 balanced records: 20,000 FOPDT and 20,000 SOPDT |
| SFT | Precision | bf16 |
| SFT | Maximum sequence | 6144 |
| SFT | Learning rate | $1\times10^{-5}$ |
| SFT | Batch configuration | 1 sequence per GPU; 8 gradient-accumulation steps |
| SFT | Regularization | Warmup ratio 0.03; weight decay 0.01; maximum gradient norm 1.0 |

| Stages | Parameters | Setting/Value |
|---|---|---|
| SFT | Validation | 5% stratified split; evaluation and checkpointing every 50 steps |
| SFT | Epoch | 3 |
| GRPO | Optimization | PI-GRPO + QLoRA |
| GRPO | Policy initialization | Selected balanced SFT model |
| GRPO | Quantization | 4 bit, NF4 |
| GRPO | LoRA rank and scaling factor | $\boldsymbol{r = 16, \alpha = 32}$ |
| GRPO | Learning rate | $1\times10^{-5}$ to $1\times10^{-6}$ |
| GRPO | Batch size per GPU | 4 |
| GRPO | Maximum training steps | 800, 1 policy epoch per rollout |
| GRPO | Group size | 4 |
| GRPO | Token limits | 6144 prompt tokens; 64 completion tokens |
| GRPO | Learning-rate schedule | Linear warmup to $1\times10^{-5}$ over 50 steps, followed by cosine decay to $1\times10^{-6}$ at step 800 |
| GRPO | Distributed sampling | 5 GPUs; 4 online prompts per GPU per step; 4 candidate completions per prompt |
| GRPO | Micro-batch size | 4 completions |
| GRPO | Training generation parameters | Temperature 0.9; top-p 0.95; top-k 50 |
| GRPO | Policy clipping | Clipping range 0.2; maximum gradient norm 1.0 |
| GRPO | Adaptive KL control | Initial coefficient 0.10; target KL 0.05; coefficient bounds [0.02,1.0] |
| GRPO | Validation protocol | Step-0 SFT baseline and a frozen set of 100 FOPDT plus 100 SOPDT cases; greedy evaluation every 50 steps |
| GRPO | Checkpoint selection | Pass@1, mean bounded IAE improvement, mean reward, and lower KL, compared lexicographically |
| SFT/GRPO | Random seeds | SFT data: 81001; SFT split: 42; PI-GRPO: 91001 |

### S5. Physics-Informed Reward Function

The reward is calculated entirely from deterministic parsing, analytical stability checks, and the shared closed-loop simulator. It does not require a learned preference model. Invalid or unsafe candidates receive a fixed penalty, whereas safe candidates are separated into unsuccessful and successful branches before their relative quality is considered. For a safe candidate, the internal quality score $Q$ combines graded stability, time-domain performance, a separate IAE-improvement term, dimensionless-gain regularization, and format compliance. The final reward then maps $Q$ to a branch fixed by safety and task completion.

$$R = \begin{cases} -1, & \text{invalid or unsafe} \\ -\frac{1}{2} + \frac{Q+1}{4}, & \text{safe but unsuccessful} \\ \frac{1}{2} + \frac{Q+1}{4}, & \text{safe and successful} \end{cases} \quad \text{(S12)}$$

$$Q = \text{clip}\left(0.35R_{stab} + 0.35R_{perf} + 0.15R_{iae} + 0.10R_{gain} + 0.05R_{fmt}, -1, 1\right) \quad \text{(S13)}$$

$R_{stab}$ is the graded Routh/phase-margin score. $R_{perf}$ combines overshoot, normalized settling time, and steady-state error without IAE. $R_{iae}$ separately measures IAE improvement. $R_{gain}$ is a bounded penalty on excessive dimensionless gains. and $R_{fmt}$ measures parseability and concise canonical output.

The reward is set to -1 when (1) parsing failure, (2) Kp<=0, Ki<0, or Kd<0, (3) a non-finite or divergent simulation, (4) Routh-Hurwitz failure, marginality, or indeterminacy, or (5) a non-positive available phase margin. A Routh-stable case with no detected unity-gain crossover remains safe but receives $R_{stab} = 0$. Among safe candidates, task success requires a settled response, overshoot strictly below 15%, and steady-state error strictly below 1%. The safe-unsuccessful and safe-successful intervals are [-0.5,0] and [0.5,1], respectively. Task success is defined by $S = 1\{\text{finite} \wedge \text{settled} \wedge M_p < 15\% \wedge e_{ss} < 1\%\}$.

### S5.1 Stability Reward

The stability term is implemented as a two-stage safety assessment. First, a closed-loop characteristic polynomial is obtained and screened by the Routh-Hurwitz criterion. Second, the phase margin is evaluated independently from the exact delayed frequency response. The Padé approximation is therefore used only for the polynomial-based Routh-Hurwitz calculation; it is not used to calculate phase margin.

For a positive-gain process, the FOPDT and SOPDT transfer functions and the ideal PID controller are defined as follows:

$$G_{\text{FOPDT}}(s) = \frac{Ke^{-\theta s}}{Ts+1}, G_{\text{SOPDT}}(s) = \frac{Ke^{-\theta s}}{(\tau_1 s+1)(\tau_2 s+1)}, C(s) = K_p + \frac{K_i}{s} + K_d s \quad \text{(S14)}$$

where $K$ is the process gain, $T$, $\tau_1$, and $\tau_2$ are time constants, $\theta$ is the dead time, and $K_p$, $K_i$, and $K_d$ are PID gains.

Under unity negative feedback, the closed-loop poles are the roots of the characteristic equation below. This equation follows from $Y(s) = C(s)G_p(s)[R(s) - Y(s)]$. Rearranging this equation gives

$$\frac{Y(s)}{R(s)} = \frac{C(s)G_p(s)}{1+C(s)G_p(s)} \quad \text{(S15)}$$

The closed-loop poles are therefore determined by the characteristic equation $1 + C(s)G_p(s) = 0$. For Routh-Hurwitz screening only, the delay is replaced by the first-order [1/1] Padé approximation. Let $d =$ $\theta/2$. Clearing the denominator after substitution yields an ordinary polynomial, , with one additional order introduced by the [1/1] Padé approximation.

$$e^{-\theta s} \approx \frac{1-\theta s/2}{1+\theta s/2} = \frac{1-ds}{1+ds}, d = \frac{\theta}{2} \tag{S16}$$

For an FOPDT process with an active integral term ($K_i > 0$), multiplying the characteristic equation by $s(1+ds)(Ts+1)$ gives the cubic polynomial below. If $\theta = 0$, $d$ is set to zero and the exact no-delay polynomial is used.

$$\Delta_F(s) = (Td - K\,d\,K_d)s^3 + [T + d + K(K_d - dK_p)]s^2 + [1 + K(K_p - dK_i)]s + KK_i = 0 \tag{S17}$$

For an SOPDT process, let $b_2 = \tau_1\tau_2$ and $b_1 = \tau_1 + \tau_2$. Multiplication by $s(1+ds)(\tau_1 s + 1)(\tau_2 s + 1)$ gives the fourth-order polynomial and coefficients below.

$$\Delta_S(s) = a_4 s^4 + a_3 s^3 + a_2 s^2 + a_1 s + a_0 = 0, a_4 = b_2 d, a_3 = b_2 + b_1 d - K\,d\,K_d, a_2 = b_1 + d + K(K_d - dK_p), a_1 = 1 + K(K_p - dK_i), a_0 = KK_i \tag{S18}$$

When $K_i$ is numerically zero, the implementation removes the controller integrator and applies the Routh construction to the reduced P/PD polynomial rather than retaining an artificial zero root. The code uses a general Routh array at the actual polynomial degree. For interpretation, the cubic condition is $a_i > 0$ together with $a_2 a_1 - a_3 a_0 > 0$; the quartic condition is $a_i > 0$ together with $a_3 a_2 - a_4 a_1 > 0$ and $a_3 a_2 a_1 - a_4 a_1^2 - a_3^2 a_0 > 0$.

$$\text{Cubic condition} = \begin{cases} a_i > 0 \\ a_2 a_1 - a_3 a_0 > 0 \end{cases} \tag{S19a}$$

$$\text{Quartic condition} = \begin{cases} a_i > 0 \\ a_3 a_2 - a_4 a_1 > 0 \\ a_3 a_2 a_1 - a_4 a_1^2 - a_3^2 a_0 > 0 \end{cases} \tag{S19b}$$

All coefficients are normalized before constructing the Routh table. A non-positive or near-zero leading coefficient, a zero first-column entry, an all-zero row, a non-finite table entry, or a first-column sign change is treated as unsafe. Thus marginal and indeterminate cases are not accepted as stable merely because a nominal symbolic condition happens to hold.

Phase margin is evaluated separately from the exact delayed loop $L(j\omega) = C(j\omega)G_0(j\omega)e^{-j\omega\theta}$. The controller phase is $\operatorname{atan2}(K_d\omega - K_i/\omega, K_p)$, and the exact delay contributes $-\omega\theta$. A logarithmic grid from $10^{-5}/T_c$ to $10^{5}/T_c$, with $T_c = T$ for FOPDT and $T_c = \tau_1 + \tau_2$ for SOPDT, identifies all unity-gain crossover brackets; each bracket is refined by bisection, and the minimum resulting margin is reported.

$$L(j\omega) = C(j\omega)G_0(j\omega)e^{-j\omega\theta},\ |L(j\omega_c)| = 1,\ PM = \min_{\omega_c}[180^\circ + \angle L(j\omega_c)] \tag{S20}$$

A Routh-stable candidate with a positive available phase margin receives a graded stability score. If no crossover is found, $R_{stab} = 0$ and the candidate remains safe. Routh failure, marginality, indeterminacy, or $PM \leq 0°$ is unsafe and activates the hard-failure reward branch.

$$R_{stab} = \begin{cases} 1 & PM > 45° \\ 0.5 + \frac{PM - 30°}{30°} & 30° < PM \leq 45° \\ \frac{PM}{60°} & 0° < PM \leq 30° \\ 0 & \text{PM unavailable and Routh safe} \\ -1 & \text{Routh unsafe/marginal/indeterminate or } PM \leq 0° \end{cases} \tag{S21}$$

This arrangement separates two purposes that should not be conflated: the Padé/Routh test supplies an algebraic stability-screening gate, whereas the exact-delay phase margin differentiates the robustness of candidates that pass that gate.

### S5.2 Performance Reward

The performance term describes response shape and completion speed. Overshoot $M_p$ and steady-state error $e_{ss}$ are expressed in percent, while settling time $T_s$ is normalized by the process-specific balanced-IMC reference settling time $t_{s,\mathrm{ref}}$, giving $q_s = \frac{T_s}{T_{s,\mathrm{ref}}}$.

$$q_s = \frac{T_s}{T_{s,\mathrm{ref}}},\ T_{s,\mathrm{ref}} = \theta - \ln(0.05)\lambda_{balanced},\ \lambda_{balanced} = 2\max(\theta, 0.1T_c) \tag{S22}$$

The corresponding bounded component rewards are:

$$r_{\mathrm{os}} = \begin{cases} 1, M_p \leq 5; \\ 1 - \frac{M_p - 5}{10}, 5 < M_p \leq 15; \\ -\frac{0.5(M_p - 15)}{15}, 15 < M_p \leq 30; \\ -0.5, M_p > 30 \end{cases} \tag{S23}$$

$$r_{\mathrm{set}} = \begin{cases} 1, & q_s \leq 1, \\ 1 - 1.3(q_s - 1), & 1 < q_s \leq 2, \\ -0.3, & q_s > 2. \end{cases} \tag{S24}$$

$$r_{\mathrm{ss}} = \begin{cases} 1, e_{ss} \leq 0.1; \\ 1 - \frac{e_{ss} - 0.1}{0.9}, 0.1 < e_{ss} \leq 1; \\ -0.5(e_{ss} - 1), 1 < e_{ss} \leq 2; \\ -0.5, e_{ss} > 2 \end{cases} \tag{S25}$$

The aggregated performance reward is

$$R_{perf} = \mathrm{clip}\left(\frac{5}{12}r_{\mathrm{os}} + \frac{4}{12}r_{\mathrm{set}} + \frac{3}{12}r_{\mathrm{ss}}, -1, 1\right) \tag{S26}$$

The piecewise terms provide interpretable engineering preferences while avoiding a single absolute settling-time threshold across processes with widely different time scales.

### S5.3 Format Reward

The format reward evaluates whether the generated completion provides a complete, concise, and machine-readable set of PID gains. Let $y$ denote the generated completion. Four indicators are used: $P(y) = 1$ if finite values of all three gains $K_p$, $K_i$, and $K_d$ can be parsed; $C(y) = 1$ if the entire completion exactly matches a canonical output format; $L(y) = 1$ if generation reaches the maximum completion length without terminating with an end-of-sequence token; and $N(y)$ denotes the number of characters after removing leading and trailing whitespace.

The two canonical output formats are: Kp=<number>, Ki=<number>, Kd=<number> and P:<number>; I:<number>; D:<number>, where = may replace : in the second format. Letter matching is case-insensitive, optional whitespace is permitted, and each gain may be represented as an integer, decimal number, or scientific-notation value.

The format reward is defined as follows:

$$R_{fmt} = \begin{cases} 0 & L(y) = 1 \\ 0 & L(y) = 0, P(y) = 0 \\ 1 & L(y) = 0, C(y) = 1 \\ 0.7 & L(y) = 0, C(y) = 0, P(y) = 1, and\ N(y) \leq 120 \\ 0.4 & L(y) = 0, C(y) = 0, P(y) = 1, and\ N(y) > 120 \end{cases} \quad \text{(S27)}$$

Here, a score of 1.0 is assigned only when the complete response exactly follows one of the canonical formats. A parseable but non-canonical response receives 0.7 when it contains no more than 120 characters and 0.4 when it is longer. This distinction discourages unnecessary explanations while still allowing physically valid PID gains to be evaluated. A completion that reaches the generation-length limit receives zero format reward even if a complete PID triplet has already been produced, because the response did not terminate normally.

The parser first searches for explicitly labelled gains using the aliases $K_p$, Kp, or P; $K_i$, Ki, or I; and $K_d$, Kd, or D. If labelled values are unavailable, a response of at most 120 characters containing between three and five numerical values is accepted as a conservative fallback, with the first three values interpreted as $K_p$, $K_i$, and $K_d$, respectively. All extracted values must be finite.

The format component has a weight of 0.05 in $Q$. It does not directly measure closed-loop quality; it ensures that a controller can be executed reliably by the simulator. If no complete PID triplet can be parsed, simulation is skipped and $R = -1$. If the completion reaches the length limit but still contains a complete triplet, the controller remains eligible for stability and performance evaluation, but Rfmt is set to zero.

**S5.4 IAE Reward**

The separate IAE term compares the new response with the current controller and maps the normalized improvement smoothly to [-1,1]:

$$R_{iae} = \tanh\left(2\rho_{(IAE)}\right), \rho_{(IAE)} = \frac{IAE_{(prev)} - IAE_{(new)}}{\max\left(\left|IAE_{prev}\right|, 10^{-8}\right)} \quad \text{(S28)}$$

**S5.5 Gain regularization**

The gain regularizer is a bounded soft penalty, not an additional stability test and not a hard PID-parameter bound. Its purpose is to reduce a training-time preference for disproportionately large proportional, integral, or derivative action while retaining process-dependent controller scaling. This is achieved by comparing dimensionless rather than raw gains.

For FOPDT, the characteristic time is $T_c = T$; for SOPDT, $T_c = \tau_1 + \tau_2$. Proportional and integral gains are scaled by the process gain and characteristic time, whereas the derivative gain is divided by the characteristic time.

$$g_p = KK_p,\ g_i = KK_iT_c,\ g_d = \frac{KK_d}{T_c}, \quad \text{(S29)}$$

The reference is calculated independently for each sampled process from its balanced IMC controller. Each dimensionless reference component is lower-bounded by 0.01. This per-process reference preserves the intended controller scale under changing gain, lag, and dead time.

$$g_{j,ref} = \max\left(0.01, g_j\left(K_{IMC}^{balanced}\right)\right),\ j \in \{p, i, d\} \quad \text{(S30)}$$

The soft limits are four times the proportional and integral reference components and six times the derivative reference component.

$$l_p = 4g_{p,\text{ref}},\ l_i = 4g_{i,\text{ref}},\ l_d = 6g_{d,\text{ref}} \quad \text{(S31)}$$

Only excess above the corresponding reference limit is penalized. With $\epsilon = 10^{-12}$, the logarithm makes the response gradual for moderate excess but increasingly discouraging as a dimensionless gain becomes much larger than its reference scale.

$$\xi_j = \max\left(0, \log\frac{g_j + \epsilon}{l_j + \epsilon}\right), R_{gain} = -\min\left(1, \sum_{j\in\{p,i,d\}} \xi_j^2\right) \in [-1,0] \quad \text{(S32)}$$

A candidate at or below all three limits receives no gain penalty. Only components above their limits contribute, through a squared logarithmic excess, to a reward between −1 and 0. The regularizer therefore discourages disproportionate gains without acting as an additional safety test.

For safe candidates, the quality-score weights are 0.35 for stability, 0.35 for performance, 0.15 for IAE improvement, 0.10 for gain regularization, and 0.05 for format compliance. $R_{gain}$ remains a soft preference: it cannot replace the Routh-Hurwitz and exact-delay phase-margin gate, and it does not impose a hidden hard gain bound.

Overall, invalid output, invalid gain signs, non-finite or divergent simulation, an unsafe or indeterminate Routh result, and a non-positive exact-delay phase margin enter the fixed unsafe branch. Safe candidates that have not yet met the settling, overshoot, and steady-state criteria occupy $[-0.5, 0]$, whereas successful candidates occupy $[0.5, 1]$.

**S6. Evaluation Protocol and Transfer-Function Pool**

The paper evaluation pool contains 100 FOPDT and 100 SOPDT cases generated with independent seeds 61001 and 62001. Process parameters are obtained by Latin hypercube sampling. Dead time lies between 1 and 200 s and is stratified by $\rho = \theta / T_c$ into low, medium, delay-dominant, and strong-delay bands with a 4:3:2:1 allocation in each group of ten cases. Each case is retained only if the initial response is finite, does not satisfy the acceptance criteria, remains within $|y| \leq 3$, and has an IAE at least 1.5 times that of the corresponding balanced IMC controller. Demonstration cases in the prompt, paper evaluation cases, GRPO validation cases, and training cases are separated by complete-case SHA-256 hashes. The complete 100-group paired pool is supplied in Section S10.

For final benchmark evaluation, all hosted and local models use identical generation settings, with a temperature of 0.1, top-p of 0.1, and reasoning or thinking mode explicitly disabled. The local Base-0.6B, SFT-0.6B, and PI-GRPO-0.6B models are served through vLLM with an inference seed of 42. No seed is specified for the hosted APIs because deterministic seed control was not available under the employed service configurations. DeepSeek-V4-Flash is accessed using the model identifier deepseek-v4-flash, whereas Qwen3.7-Plus is accessed using the versioned identifier qwen3.7-plus-2026-05-26. These settings are held fixed across process classes, prompt variants, and tuning rounds.

**S7. Ablation Study on Prompt Strategies**

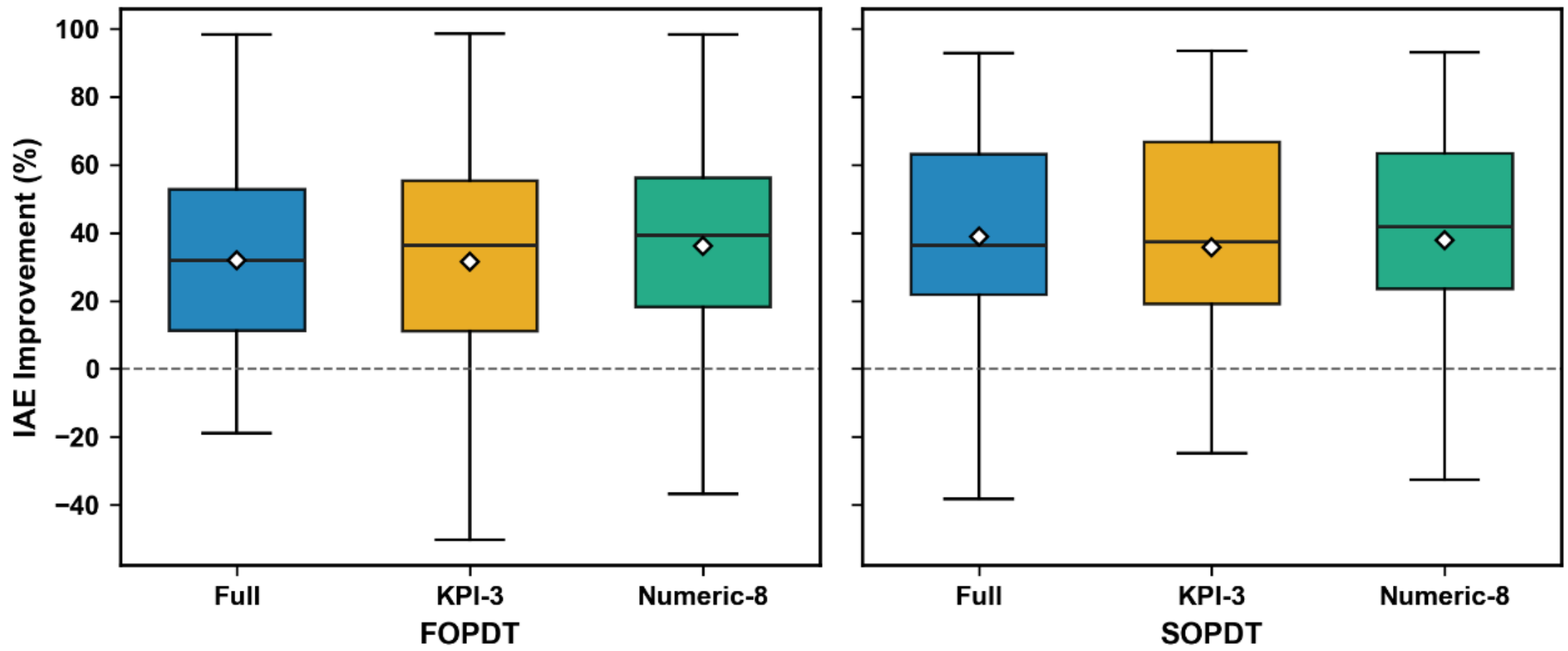

**Figure S1.** Distribution of IAE improvement under the Full, KPI-3, and Numeric-8 prompt strategies for DeepSeek-V4-Flash.

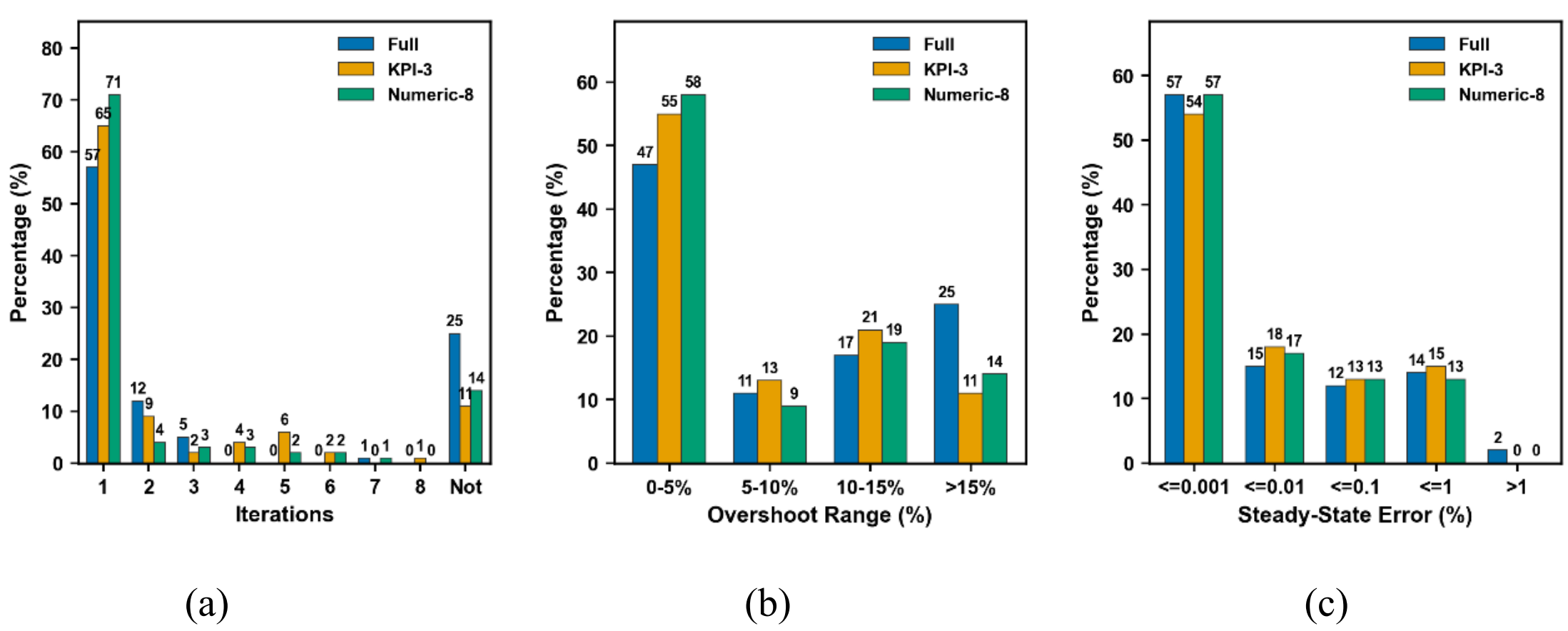

(a) (b) (c)

**Figure S2.** Batch results by DeepSeek-V4-Flash for 100 FOPDT cases: (a) iteration at which the first acceptable controller is obtained, where "Not" denotes failure within eight rounds; (b) overshoot distribution after the final tuning round; and (c) steady-state-error distribution after the final tuning round.

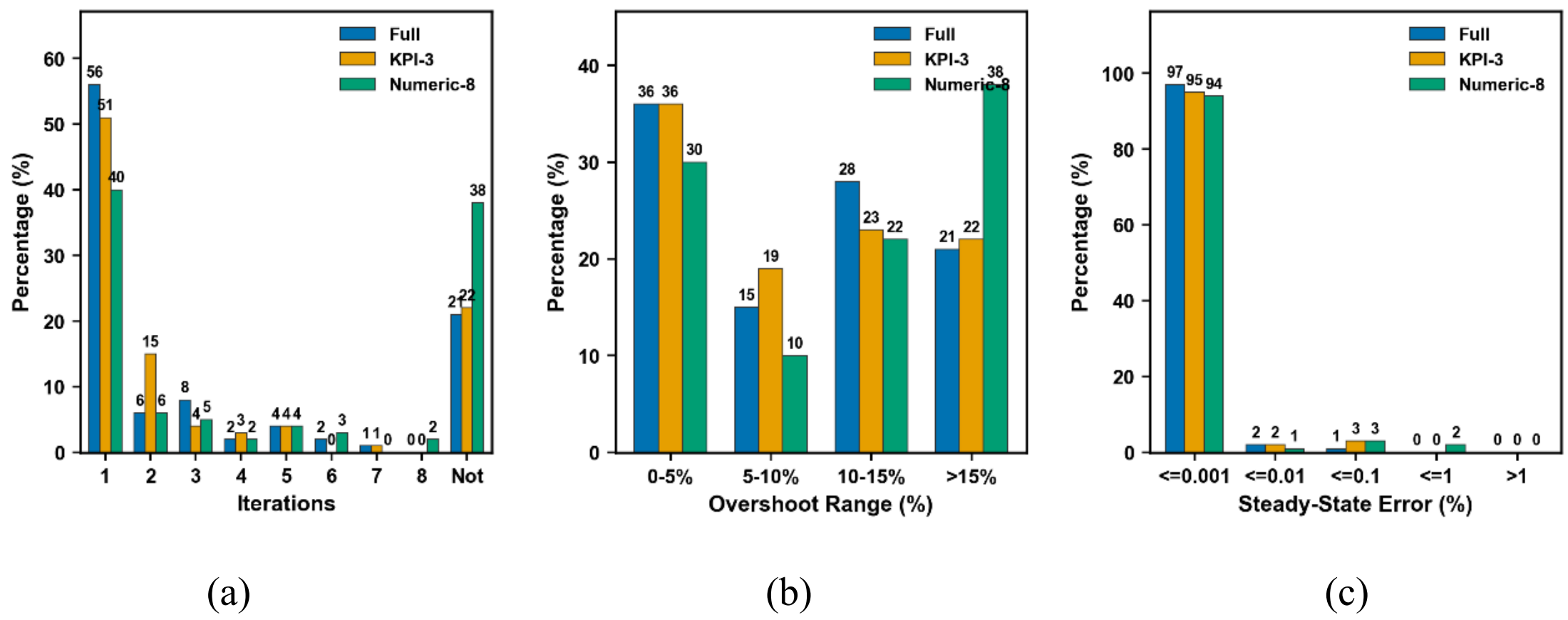

(a) (b) (c)

**Figure S3.** Batch results by DeepSeek-V4-Flash for 100 SOPDT cases: (a) iteration at which the first acceptable controller is obtained, where "Not" denotes failure within eight rounds; (b) overshoot distribution after the final tuning round; and (c) steady-state-error distribution after the final tuning round.

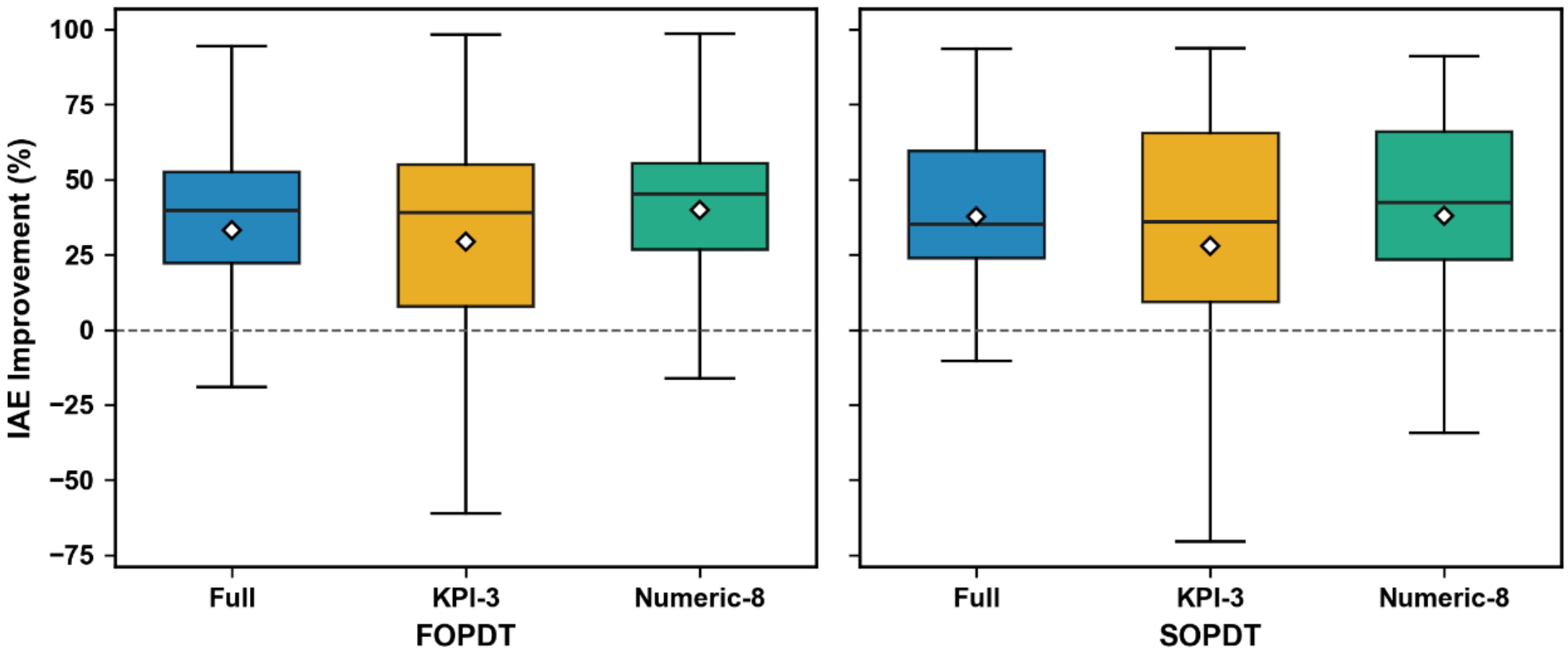

**Figure S4.** Distribution of IAE improvement under the Full, KPI-3, and Numeric-8 prompt strategies for Qwen3.7-Plus.

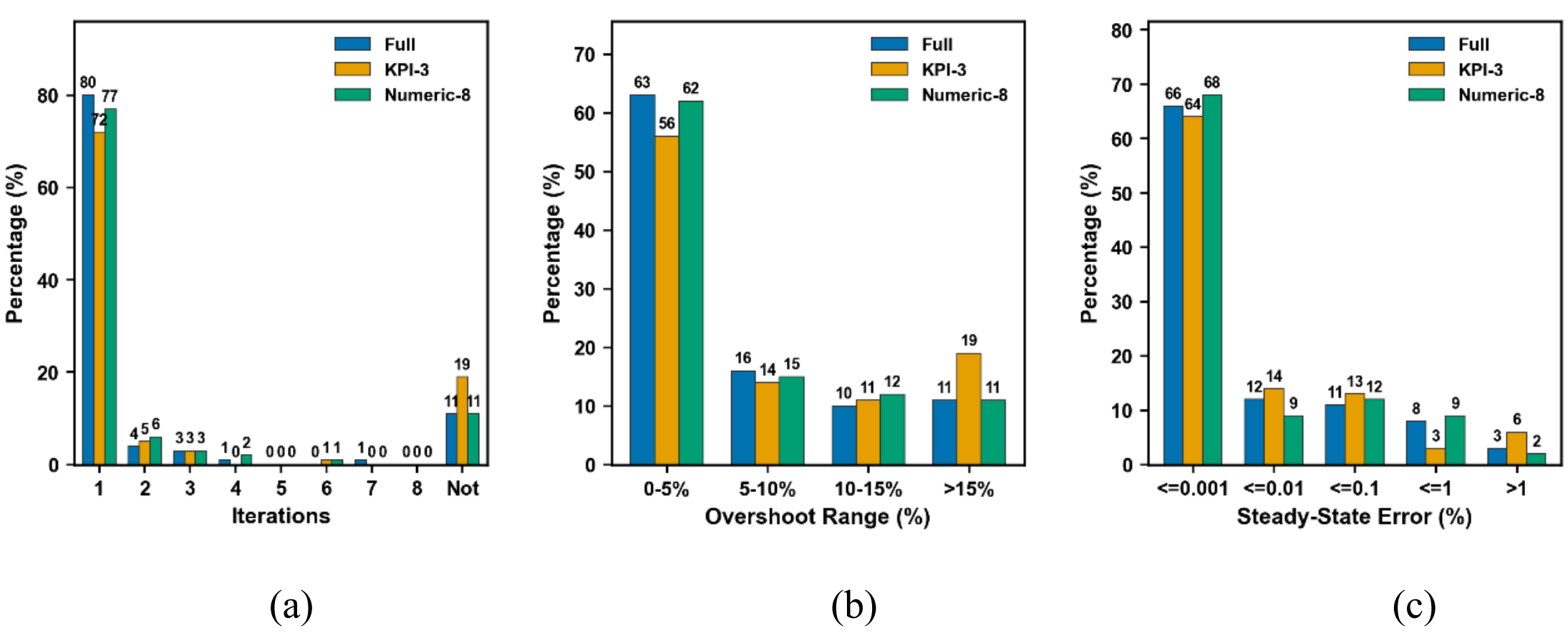

(a) (b) (c)

**Figure S5.** Batch results by Qwen3.7-Plus for 100 FOPDT cases: (a) iteration at which the first acceptable controller is obtained, where “Not” denotes failure within eight rounds; (b) overshoot distribution after the final tuning round; and (c) steady-state-error distribution after the final tuning round.

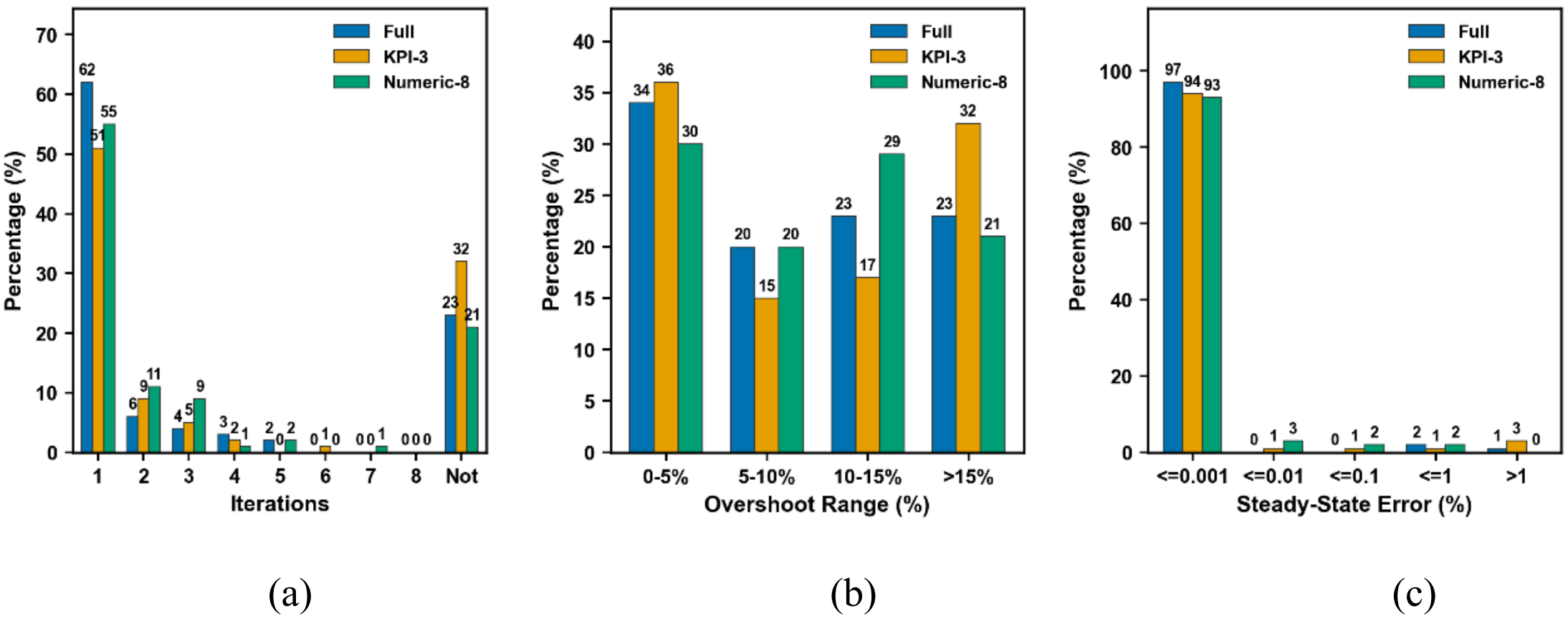

(a) (b) (c)

**Figure S6.** Batch results by Qwen3.7-Plus for 100 SOPDT cases: (a) iteration at which the first acceptable controller is obtained, where “Not” denotes failure within eight rounds; (b) overshoot distribution after the final tuning round; and (c) steady-state-error distribution after the final tuning round.

## S8. Detailed Statistics by SFT-0.6B and GRPO-0.6B

On FOPDT processes, Pass@1 is 13% for Base-0.6B, 94% after SFT, and 92% after PI-GRPO. The SFT model therefore already resolves most first-order cases, and reinforcement learning does not increase their success rate. It does, however, reduce mean overshoot from 2.99% to 0.96%, revealing a shift toward more conservative first-step actions.

The benefit of PI-GRPO is concentrated in the more difficult SOPDT cohort. Pass@1 rises from 18% for Base-0.6B to 79% after SFT and 96% after PI-GRPO, while mean overshoot falls from 9.89% for SFT to 4.84% for PI-GRPO. The adapted 0.6B model also exceeds the one-step success of DeepSeek-V4-Flash (56%) and Qwen3.7-Plus (62%) under the same test protocol, showing that task-specific training can compensate for much of the general-model scale gap.

This gain in first-step reliability is accompanied by a clear performance trade-off. Mean first-round IAE increases from 193.1 to 231.2 for FOPDT and from 81.4 to 107.2 for SOPDT; across 200 paired cases, PI-GRPO improves IAE in 75 and worsens it in 125. The learned policy largely preserves proportional and derivative action while reducing integral action, which suppresses overshoot but can slow error removal or leave residual error in a small subset of FOPDT cases. PI-GRPO should therefore be interpreted as a stability-first refinement, not as universal IAE minimization.

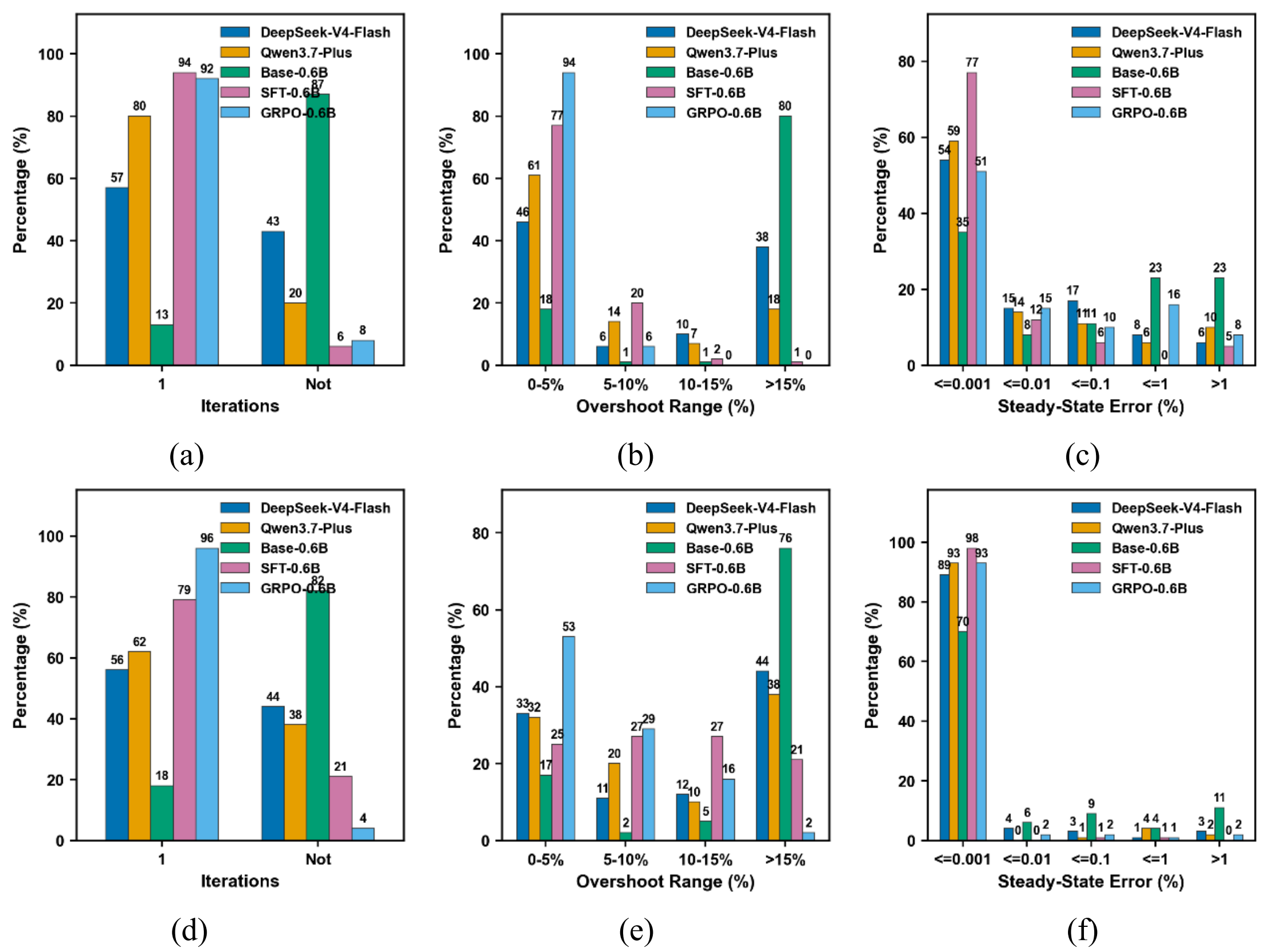

**Figure S7.** One-step benchmark across DeepSeek-V4-Flash, Qwen3.7-Plus, Base-0.6B, SFT-0.6B, and GRPO-0.6B for (a-c) FOPDT and (d-f) SOPDT cases: tuning success, overshoot, and steady-state error.

## S9. Response Curves by SFT-0.6B and GRPO-0.6B

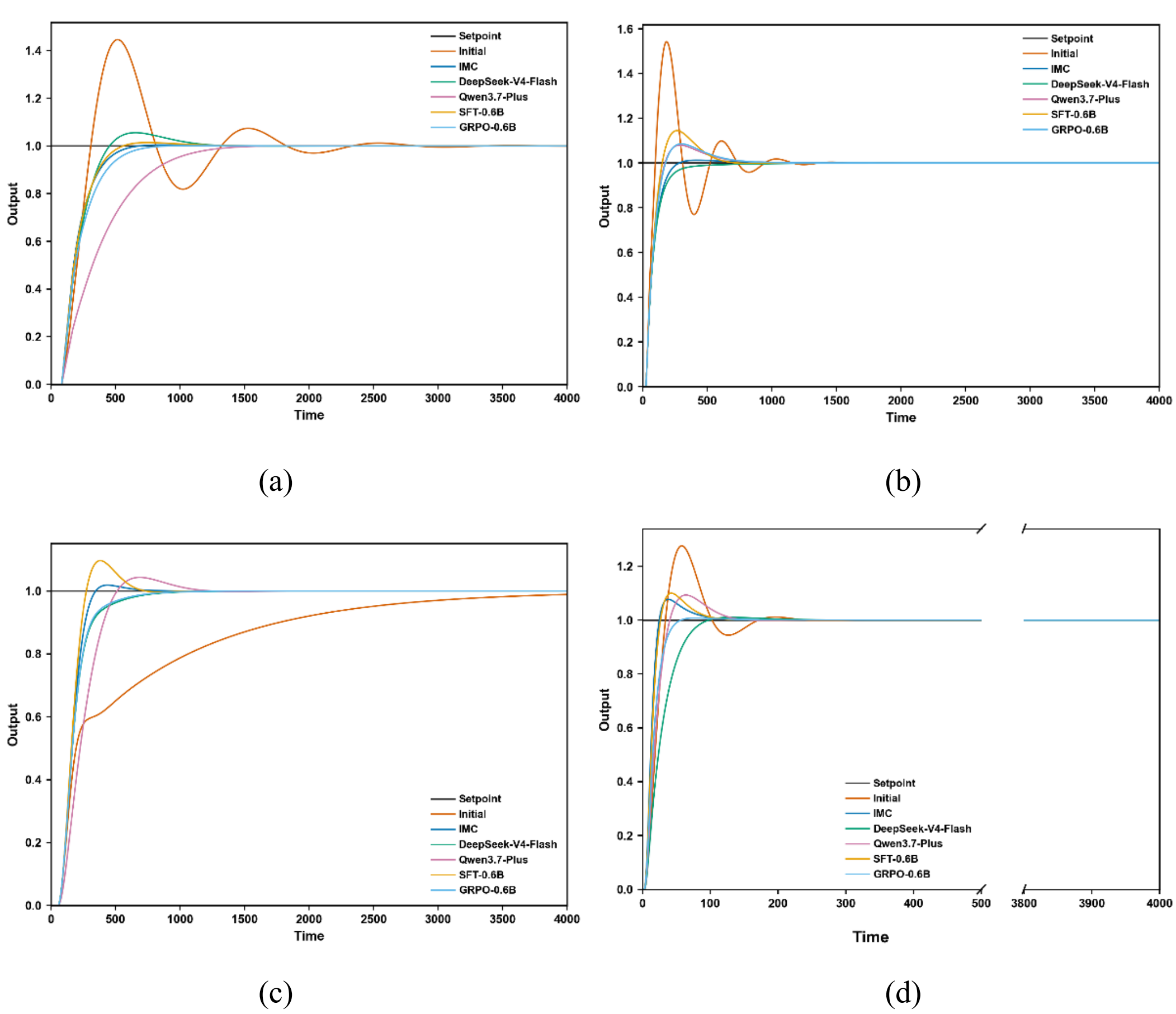

**Figure S8.** Step-response comparison for (a) $G(s) = \frac{0.667}{204.269s+1}e^{-84.060s}$, (b) $G(s) = \frac{0.839}{336.266s+1}e^{-26.350s}$, (c) $G(s) = \frac{0.418e^{-55.630s}}{(54.645s+1)(76.887s+1)}$, and (d) $G(s) = \frac{1.620e^{-2.903s}}{(6.257s+1)(26.231s+1)}$ under the initial controller, IMC, DeepSeek-V4-Flash, Qwen3.7-Plus, SFT-0.6B, and GRPO-0.6B.

**Table S2.** PID settings and time-domain metrics for the representative FOPDT and SOPDT cases.

| Case | Method | $K_p$ | $K_i$ | $K_d$ | IAE | $M_p$ (%) | $e_{ss}$ (%) | Rounds |
|---|---|---|---|---|---|---|---|---|
| C1 | Initial | 0.845 | 0.015 | 61.287 | 437.56 | 44.58 | 0.06 | - |
| C1 | IMC | 1.758 | 0.007 | 61.287 | 214.58 | 0.43 | 0.00 | - |
| C1 | DeepSeek-V4-Flash | 1.500 | 0.008 | 61.287 | 237.16 | 5.56 | 0.00 | 1 |
| C1 | Qwen3.7-Plus | 0.845 | 0.004 | 61.287 | 401.63 | 0.10 | 0.00 | 1 |
| C1 | Qwen3-0.6B-Base[#] | 0.845 | 0.015 | 61.287 | 441.07 | 45.16 | 0.08 | 8 |
| C1 | SFT-0.6B | 1.700 | 0.007 | 61.287 | 217.27 | 1.44 | 0.00 | 1 |
| C1 | GRPO-0.6B | 1.600 | 0.006 | 61.287 | 236.33 | 0.161 | 0.00 | 1 |
| C2 | Initial | 5.179 | 0.120 | 186.774 | 185.45 | 54.16 | 0.00 | - |

| Case | Method | $K_p$ | $K_i$ | $K_d$ | IAE | $M_p$ (%) | $e_{ss}$ (%) | Rounds |
|---|---|---|---|---|---|---|---|---|
| C2 | IMC | 5.179 | 0.015 | 65.660 | 92.65 | 1.20 | 0.00 | - |
| C2 | DeepSeek-V4-Flash | 5.179 | 0.012 | 60.000 | 99.43 | 0.00 | 0.00 | 1 |
| C2 | Qwen3.7-Plus | 5.179 | 0.024 | 58.730 | 102.37 | 7.93 | 0.00 | 1 |
| C2 | Qwen3-0.6B-Base | 1.451 | 0.003 | 96.702 | 336.60 | 0.02 | 0.00 | 1 |
| C2 | SFT-0.6B | 5.179 | 0.032 | 83.088 | 111.63 | 14.423 | 0.00 | 1 |
| C2 | GRPO-0.6B | 5.179 | 0.024 | 75.877 | 106.03 | 8.500 | 0.00 | 1 |
| C3 | Initial | 1.885 | 0.004 | 60.202 | 625.89 | 0.00 | 1.11 | - |
| C3 | IMC | 1.885 | 0.014 | 60.202 | 176.05 | 1.89 | 0.00 | - |
| C3 | DeepSeek-V4-Flash | 1.885 | 0.012 | 60.202 | 199.38 | 0.00 | 0.00 | 1 |
| C3 | Qwen3.7-Plus | 0.943 | 0.011 | 60.202 | 260.15 | 4.35 | 0.00 | 2 |
| C3 | Qwen3-0.6B-Base[#] | 1.885 | 0.014 | 60.202 | 590.17 | 0.0 | 0.83 | 1 |
| C3 | SFT-0.6B | 1.885 | 0.017 | 60.202 | 185.05 | 9.670 | 0.00 | 1 |
| C3 | GRPO-0.6B | 1.885 | 0.012 | 60.202 | 196.11 | 0.000 | 0.00 | 1 |
| C4 | Initial | 0.913 | 0.066 | 10.776 | 33.12 | 27.57 | 0.00 | - |
| C4 | IMC | 2.133 | 0.066 | 10.776 | 16.20 | 7.74 | 0.00 | - |
| C4 | DeepSeek-V4-Flash | 0.913 | 0.022 | 10.776 | 29.94 | 1.02 | 0.00 | 1 |
| C4 | Qwen3.7-Plus | 1.250 | 0.045 | 10.776 | 23.71 | 9.32 | 0.00 | 1 |
| C4 | Qwen3-0.6B-Base | 1.359 | 0.021 | 19.207 | 29.70 | 0.00 | 0.00 | 1 |
| C4 | SFT-0.6B | 1.900 | 0.066 | 10.776 | 17.95 | 10.006 | 0.00 | 1 |
| C4 | GRPO-0.6B | 1.600 | 0.036 | 10.776 | 18.27 | 0.772 | 0.00 | 1 |

[#]Minor IAE differences arise because the initial PID values retain full precision, whereas the reported and model-returned values are rounded to three decimal places.

### S10. FOPDT and SOPDT Benchmarks

**Table S3.** FOPDT and SOPDT benchmarks.

| | FOPDT | | | SOPDT | | | |
|---|---|---|---|---|---|---|---|
| group | $K$ | $T$ | $\theta$ | $K$ | $\tau_1$ | $\tau_2$ | $\theta$ |
| 1 | 0.257886 | 109.770 | 77.91 | 2.09872 | 58.419 | 70.7406 | 91.31 |
| 2 | 0.740899 | 205.893 | 125.1 | 0.203089 | 68.3982 | 88.0125 | 81.95 |
| 3 | 0.740899 | 205.893 | 125.1 | 0.856887 | 41.0117 | 60.9337 | 65.33 |
| 4 | 0.396741 | 321.396 | 191.3 | 0.108402 | 65.0308 | 73.907 | 110.8 |
| 5 | 0.680716 | 328.687 | 174.9 | 1.78544 | 7.79252 | 77.297 | 47.09 |
| 6 | 0.386363 | 253.634 | 151.7 | 1.36869 | 65.3915 | 81.2306 | 114.5 |
| 7 | 0.599392 | 240.358 | 129.7 | 1.02 | 23.8656 | 78.9524 | 74.25 |
| 8 | 0.334481 | 129.337 | 95.39 | 2.19784 | 10.499 | 66.7488 | 56.51 |
| 9 | 0.340705 | 382.178 | 199.5 | 0.203089 | 68.3982 | 88.0125 | 81.95 |
| 10 | 0.768169 | 237.928 | 181.7 | 2.37422 | 10.4578 | 41.0343 | 34.29 |
| 11 | 0.39485 | 518.868 | 172.3 | 2.26145 | 26.889 | 34.4671 | 22.17 |
| 12 | 0.768169 | 237.928 | 110.3 | 1.12352 | 18.0693 | 61.9562 | 25.83 |
| 13 | 0.268322 | 293.766 | 116.8 | 2.33429 | 11.1671 | 77.5251 | 18.02 |
| 14 | 0.85776 | 134.182 | 29.91 | 0.418177 | 54.6448 | 76.8869 | 55.63 |
| 15 | 0.337788 | 413.635 | 158.9 | 0.418177 | 54.6448 | 76.8869 | 55.63 |
| 16 | 0.730958 | 227.41 | 59.6 | 0.156643 | 32.0018 | 33.3922 | 14.35 |
| 17 | 0.291143 | 279.265 | 120.1 | 2.24474 | 22.143 | 86.0608 | 53.43 |
| 18 | 0.666602 | 204.269 | 84.06 | 1.76775 | 35.2087 | 74.0123 | 25.91 |
| 19 | 0.452147 | 561.815 | 179.1 | 0.11827 | 46.2327 | 48.8767 | 42.18 |
| 20 | 0.636685 | 526.373 | 158.1 | 2.64156 | 69.8024 | 81.4038 | 67.7 |
| 21 | 0.468158 | 191.951 | 37.47 | 2.19992 | 12.126 | 83.4347 | 15.35 |
| 22 | 0.848814 | 566.797 | 69.7 | 2.19792 | 62.9927 | 68.2599 | 15.45 |
| 23 | 0.303865 | 209.345 | 12.65 | 1.16484 | 73.0695 | 74.3913 | 16.96 |
| 24 | 0.825768 | 211.511 | 19.41 | 1.22479 | 45.4086 | 63.9628 | 8.317 |
| 25 | 0.691086 | 320.448 | 28.57 | 1.22479 | 45.4086 | 63.9628 | 8.317 |
| 26 | 0.694829 | 143.153 | 28.27 | 1.45065 | 24.0769 | 29.3186 | 3.832 |
| 27 | 0.473576 | 312.129 | 48.59 | 2.86454 | 39.7143 | 47.4648 | 13.04 |
| 28 | 0.371903 | 346.938 | 18.31 | 2.07853 | 15.7961 | 35.5969 | 6.925 |
| 29 | 0.494355 | 533.821 | 103.7 | 0.145208 | 60.4211 | 84.5757 | 9.487 |
| 30 | 0.858456 | 426.532 | 30.7 | 1.28127 | 37.5644 | 42.4189 | 8.072 |
| 31 | 0.818987 | 257.881 | 118.4 | 1.08734 | 47.5183 | 53.0585 | 41.46 |
| 32 | 0.309886 | 123.618 | 54.94 | 0.809952 | 13.7245 | 17.4565 | 11.35 |
| 33 | 0.675075 | 208.372 | 73.62 | 0.883576 | 3.49286 | 34.0582 | 17.38 |
| 34 | 0.62461 | 413.092 | 100.3 | 0.352111 | 42.7887 | 64.0671 | 28.61 |
| 35 | 0.823091 | 313.11 | 83.29 | 0.271988 | 38.9633 | 74.2794 | 30.98 |
| 36 | 0.738136 | 421.95 | 140.6 | 2.31427 | 52.9721 | 70.4273 | 58.05 |
| 37 | 0.64196 | 163.085 | 58.55 | 2.29633 | 37.2933 | 59.8128 | 36.35 |
| 38 | 0.356538 | 128.67 | 52.42 | 0.595455 | 7.5434 | 28.9597 | 8.37 |
| 39 | 0.305957 | 324.181 | 115.8 | 0.386634 | 21.5226 | 26.6059 | 14.41 |
| 40 | 0.76316 | 553.375 | 139.6 | 2.47196 | 13.6753 | 38.7432 | 21.97 |
| 41 | 0.219297 | 416.004 | 36.99 | 2.13552 | 9.66535 | 12.5377 | 4.235 |
| 42 | 0.808839 | 583.284 | 34.11 | 2.11516 | 16.7696 | 87.7827 | 9.089 |
| 43 | 0.678344 | 442.781 | 79.93 | 2.60949 | 10.0946 | 59.1738 | 4.527 |
| 44 | 0.774266 | 377.347 | 61.16 | 2.67908 | 45.1502 | 89.7144 | 14.6 |
| 45 | 0.379834 | 440.478 | 41 | 1.50097 | 24.7923 | 48.7552 | 14.17 |
| 46 | 0.512655 | 269.325 | 40.35 | 1.85404 | 4.58795 | 11.1899 | 3.028 |
| 47 | 0.481432 | 258.13 | 14.46 | 2.88895 | 8.69628 | 56.5091 | 6.4 |
| 48 | 0.86485 | 241.487 | 40.37 | 1.28688 | 12.823 | 71.1033 | 5.416 |
| 49 | 0.442461 | 506.421 | 80.11 | 2.37393 | 4.22943 | 18.4952 | 4.166 |
| 50 | 0.734519 | 175.794 | 21.15 | 1.2316 | 6.80854 | 45.3008 | 8.932 |
| 51 | 0.449045 | 500.551 | 37.34 | 2.67908 | 45.1502 | 89.7144 | 14.6 |
| 52 | 0.717023 | 111.98 | 13.19 | 1.62031 | 6.25732 | 26.2308 | 2.903 |
| 53 | 0.222111 | 586.406 | 72.18 | 2.57569 | 21.6372 | 61.1357 | 4.174 |
| 54 | 0.885443 | 356.236 | 67.55 | 2.93553 | 31.5828 | 78.9184 | 10.45 |
| 55 | 0.838922 | 336.266 | 26.35 | 0.525845 | 25.8376 | 35.5866 | 10.48 |
| 56 | 0.567972 | 219.911 | 42.01 | 1.53649 | 21.2137 | 33.2728 | 3.26 |
| 57 | 0.498296 | 204.151 | 32.6 | 0.773157 | 8.18262 | 19.0725 | 4.64 |
| 58 | 0.209449 | 316.749 | 50.31 | 0.976611 | 37.4653 | 40.8353 | 13.01 |
| 59 | 0.494801 | 281.587 | 14.11 | 1.51921 | 5.30603 | 45.3671 | 7.581 |
| 60 | 0.212748 | 112.593 | 19.42 | 1.87274 | 5.37207 | 27.6946 | 6.419 |
| 61 | 0.729317 | 163.813 | 2.991 | 2.93553 | 31.5828 | 78.9184 | 2.344 |
| 62 | 0.721461 | 538.968 | 22.99 | 1.42835 | 28.5408 | 32.5556 | 1.181 |
| 63 | 0.433427 | 307.126 | 11.72 | 0.61277 | 50.2826 | 75.4264 | 3.525 |
| 64 | 0.670478 | 499.644 | 1.383 | 0.61277 | 50.2826 | 75.4264 | 3.525 |
| 65 | 0.64517 | 227.877 | 7.53 | 0.966044 | 40.2383 | 80.1862 | 5.174 |
| 66 | 0.283578 | 259.162 | 7.827 | 2.48918 | 29.8185 | 45.2929 | 2.493 |
| 67 | 0.342276 | 397.094 | 2.8 | 0.611321 | 32.7383 | 76.251 | 3.442 |
| 68 | 0.283578 | 259.162 | 7.827 | 1.72781 | 51.7553 | 74.8739 | 1.689 |
| 69 | 0.701424 | 133.831 | 1.45 | 1.75286 | 27.0999 | 71.8546 | 3.718 |
| 70 | 0.827445 | 443.035 | 8.467 | 0.385006 | 9.09277 | 61.4363 | 1.876 |
| 71 | 0.511293 | 369.384 | 10.48 | 2.94507 | 18.7877 | 34.4639 | 1.093 |
| 72 | 0.389479 | 486.945 | 14.93 | 2.37934 | 20.5463 | 85.4687 | 3.156 |
| 73 | 0.828303 | 203.953 | 8.854 | 1.27814 | 10.225 | 37.4346 | 1.838 |
| 74 | 0.314365 | 271.996 | 5.557 | 2.98837 | 49.3765 | 77.7928 | 2.215 |
| 75 | 0.314365 | 271.996 | 5.557 | 1.5553 | 42.4617 | 43.4921 | 1.77 |
| 76 | 0.888628 | 171.475 | 7.489 | 1.52755 | 47.8599 | 69.193 | 4.557 |
| 77 | 0.841217 | 472.503 | 15.6 | 2.29431 | 7.66934 | 65.3397 | 3.497 |
| 78 | 0.293203 | 454.022 | 3.869 | 2.85455 | 26.0977 | 28.1929 | 2.668 |
| 79 | 0.534206 | 268.098 | 1.625 | 0.882243 | 13.5351 | 17.4383 | 1.497 |
| 80 | 0.59207 | 523.155 | 11.55 | 0.882243 | 13.5351 | 17.4383 | 1.497 |
| 81 | 0.858991 | 523.31 | 22.54 | 1.75855 | 39.9123 | 49.6078 | 2.357 |
| 82 | 0.465935 | 106.269 | 2.187 | 0.257198 | 52.359 | 70.1954 | 3.584 |
| 83 | 0.858991 | 523.31 | 22.54 | 2.94423 | 20.9497 | 39.5234 | 1.646 |
| 84 | 0.341113 | 396.145 | 6.955 | 1.1243 | 6.85013 | 14.6259 | 1.033 |
| 85 | 0.478956 | 451.286 | 7.146 | 1.02664 | 23.3845 | 55.5356 | 1.655 |
| 86 | 0.209993 | 598.523 | 19.33 | 0.431013 | 11.5704 | 17.5856 | 1.178 |
| 87 | 0.763921 | 430.741 | 6.961 | 0.255907 | 46.6152 | 81.1512 | 1.387 |
| 88 | 0.471007 | 401.954 | 17.66 | 1.25163 | 27.2628 | 33.7308 | 1.732 |
| 89 | 0.541711 | 172.615 | 3.087 | 1.12602 | 25.6497 | 28.3452 | 1.72 |
| 90 | 0.879125 | 136.526 | 1.162 | 1.95667 | 14.0942 | 70.4403 | 3.607 |
| 91 | 0.552894 | 156.755 | 1.609 | 2.28348 | 48.4435 | 80.7676 | 5.781 |
| 92 | 0.7871 | 162.054 | 5.691 | 2.8028 | 36.6736 | 51.6671 | 4 |
| 93 | 0.880154 | 406.645 | 4.448 | 1.0582 | 38.8069 | 62.719 | 4.226 |
| 94 | 0.575049 | 365.193 | 10.58 | 2.99445 | 17.2019 | 17.7682 | 1.627 |
| 95 | 0.443524 | 118.324 | 2.322 | 0.586776 | 8.63122 | 51.569 | 2.718 |
| 96 | 0.603948 | 286.696 | 12.69 | 0.324844 | 14.7765 | 49.4621 | 2.208 |
| 97 | 0.77937 | 488.878 | 16.21 | 2.0967 | 30.041 | 40.1542 | 1.947 |
| 98 | 0.256563 | 173.335 | 3.399 | 2.90279 | 20.481 | 30.1778 | 1.074 |
| 99 | 0.425097 | 291.022 | 14.25 | 2.37055 | 15.73 | 37.4497 | 2.252 |
| 100 | 0.777384 | 594.169 | 2.403 | 1.51506 | 16.7493 | 35.0997 | 1.638 |